\documentclass[%
 reprint,showkeys,
 amsmath,amssymb,
 aps,
]{revtex4-2}

\usepackage{graphicx}
\usepackage{dcolumn}
\usepackage{bm}
\usepackage{longtable}

\begin{document}

\preprint{APS/123-QED}

\title{CROSS-SECTIONS OF PHOTONEUTRON REACTIONS \\ ON $^{181}$Ta AT $E_{\rm \gamma max}$ UP TO 95 MeV } 

\author{O.S. Deiev, I.S. Timchenko, S.N. Olejnik, \\
	V.A. Kushnir, V.V. Mytrochenko, S.A. Perezhogin}


\affiliation{  NSC Kharkov Institute of Physics and Technology, National Academy of Sciences of Ukraine }%

\date{\today}

\begin{abstract}
The total bremsstrahlung flux-averaged cross-sections $\langle{\sigma(E_{\rm{\gamma max}})}\rangle$ for the photonuclear reactions $^{181}\rm{Ta}(\gamma,\textit{x}n; \textit{x} \leq 8)^{181-\textit{x}}\rm{Ta}$ have been measured in the range of end-point energies $E_{\rm \gamma max}$ up to 95~MeV. The experiments were performed with the beam from the NSC KIPT electron linear accelerator LUE-40 with the use of the activation and off-line $\gamma$-ray spectrometric technique. The calculation of average cross-sections was carried out using the cross-section values computed with the TALYS1.95 code for different level density models $LD$~1-6. A comparison between the experimental total cross-sections $\langle{\sigma(E_{\rm{\gamma max}})}\rangle$ and the theoretical values has shown satisfactory agreement for the $^{181}\rm{Ta}(\gamma,\textit{x}n)$ reactions with the escape of 1-4, 6 and 7 neutrons. The closest agreement with the measured data is observed with the $LD$5 computation version, which represents the microscopic level density model taking into account Hilaire’s combinatorial tables. 
\end{abstract}

\maketitle


\section{\label{sec:INTRO} INTRODUCTION }

Studies into multiparticle photonuclear reactions in the energy range above the giant dipole resonance (GDR) and up to the pion production threshold ($E_{\rm th} \approx 145$~MeV) are of particular interest. This is because in the energy region under consideration the change in the mechanism of photonuclear interaction takes place, i.e., detailed knowledge can be obtained about the competition between two mechanisms of nuclear photodisintegration, namely, through the GDR excitation, and the quasideuteron photoabsorption \cite{1}.

To describe the mechanism of multiparticle photonuclear reactions, various theoretical models have been developed. For example, a combined model of photonucleon reactions was worked out in Ref.~\cite{2}, which united the semiempirical model of oscillations, the quasideuteron model of photoabsorption, the exciton and evaporation models. For illustration, that model was tested using the data on relative yields from the $^{181}\rm{Ta}(\gamma,\textit{x}n)^{181-\textit{x}}\rm{Ta}$ reactions with emission of up to 6 neutrons, and also, the reactions $^{181}\rm{Ta}(\gamma,p)^{180m}\rm{Hf}$ and $^{181}\rm{Ta}(\gamma,pn)^{179m}\rm{Hf}$  with the charged particle in the outlet channel \cite{3}. The combined-model calculations were also compared with the yields of isotopes produced in the photonuclear reactions on the natural molybdenum at the end-point energy of the bremsstrahlung spectrum $E_{\rm \gamma max}$ = 67.7~MeV \cite{4}.
Rodrigues et al.~\cite{5} have proposed the Monte Carlo multicollisional intranuclear cascade model to study photonuclear reactions at intermediate energies ($20 \leq E \leq 140$~MeV); the model predictions were compared with the experimental total photoabsorption cross-sections for the nuclei Sn, Ce, Ta, and Pb (see Ref.~\cite{6}) and with the calculated data from Ref.~\cite{7} in the energy range $ E = 25$ to 132~MeV.

The modern Hauser-Feshbach nuclear reaction codes such as EMPIRE \cite{8}, TALYS \cite{9}, CCONE \cite{10}, and CoH3 \cite{11}, used to calculate photo-induced reaction cross-sections, also call for testing in a wide range of atomic masses and energies. Although at present there is a sufficient dataset on the experimental cross-sections in the GDR energy region \cite{12,13}, the use of the data on single- or two-particle reactions only, precludes testing various theoretical concepts, because the results of calculations by different models (e.g., nuclear level density $LD$ 1-6 computation in the TALYS1.95 code) differ from each other only slightly. With the increase in the number of particles in the outlet channel of the reaction, the distinctions between different theoretical models become more prominent. This enables one to choose the optimum version of the computational model using the reaction cross-section data with $x \geq 3$. However, the experimental cross-sections for photonuclear reactions with emission of a large number of particles are thus far still lacking.

The multiparticle photonuclear reactions that have relatively low cross-sections can be observed with the availability of intense incident $\gamma$-quantum fluxes. These fluxes can be provided by electron linear accelerators with the use of targets-converters for bremsstrahlung generation. It should be noted that the experiments on bremsstrahlung beams substantially complicate the procedure of determining the photonuclear reaction cross-sections. First of all, it is necessary to determine precisely the $\gamma$-quantum flux density to comply with the real experimental conditions. To this end, modern computational codes (e.g., GEANT4 or MCNP) should be used. Besides, the experiment implies the measurement of the integral characteristics of the reactions, and this calls for additional mathematical processing of the results. And yet, despite the arising difficulties, the bremsstrahlung beams are an important tool in the studies of photonuclear reactions.

Tantalum is belonging to structural, shielding, and bremsstrahlung target materials. The natural tantalum consists of a mixture of two isotopes $^{181}$Ta and $^{180\rm m}$Ta in the ratio of 8300 to 1. The $^{181}$Ta presents a heavily deformed nucleus with the quadrupole deformation parameter $\beta$ = 0.26 \cite{12}. The deformation can give rise, for example, to a complex structure in the energy dependence of the  $^{181}\rm{Ta}(\gamma,n)^{180}\rm{Ta}$ reaction cross-section, which exhibits two peaks in the GDR region. The $^{180\rm m}$Ta isotope is the only stable isomer (within the sensitivity of current techniques), which comes under the so-called bypassed nuclei, the existence of which in nature poses the problem of search/investigation of the processes of heavy nuclei formation, which are not associated with the neutron capture. 

The photodisintegration of $^{181}$Ta in the GDR range was experimentally investigated using the beams of quasimonochromatic and bremsstrahlung $\gamma$-quanta \cite{6,14,15,16,17,18,19,20}. The result was that the total photoabsorption cross-section values for $\sigma(\gamma,abs)$ and $\sigma(\gamma,s\rm n)$ were determined. The cross-sections for photoneutron reactions on $^{181}$Ta were obtained by the method of direct neutron registration in the GDR region for $(\gamma,\rm{n})$ and $(\gamma,2\rm n)$ (see Ref.~\cite{15}). In \cite{16}, studies were made into the cross-sections for the $(\gamma,\rm n)$, $(\gamma,2\rm n)$ and $(\gamma,3\rm n)$ reactions, and also, for the $(\gamma,4\rm n)$ reaction up to an energy of 36 MeV. Paper \cite{21} presents the experimental results from NEW Subaru studies for the $^{181}\rm{Ta}(\gamma,\textit{x}n; \textit{x} \leq 4)$  reaction, and draws a comparison between the data obtained at different laboratories. The observed difference in the absolute values of the reaction cross-sections determined by different methods gives impetus to conducting additional studies on tantalum photodisintegration. Note that the cross-sections for photoproton and photoneutron reactions with $x > 4$ on the $^{181}$Ta nucleus have not been measured \cite{13}.

Several photonuclear reaction studies were carried out on the $^{181}$Ta nucleus with the use of $\gamma$-bremsstrahlung beams at energies higher than the GDR energy. The experiments have given the weighted average yield values at end-point energies of the bremsstrahlung spectrum $E_{\rm \gamma max}$ = 40 and 55~MeV \cite{22}, and the relative yields at $E_{\rm \gamma max}$ = 67.7~MeV \cite{3}. 

In work \cite{23}, the bremsstrahlung flux-averaged cross-sections $\langle{\sigma(E_{\rm{\gamma max}})}\rangle$ have been measured for photoneutron reactions on $^{181}$Ta with emission of up to 8 neutrons at $E_{\rm{\gamma max}}$ = 80 to 95 MeV. Along with that, a comparison with the TALYS1.9-based computations has been made. Note that the cross-section maxima for these reactions lie at energies below 80 MeV.  The $\langle{\sigma(E_{\rm{\gamma max}})}\rangle$ data extension towards lower $E_{\rm{\gamma max}}$ will make it possible to test the calculations in the region of cross-section maxima for the $x \leq 8\rm n$ reactions, just as this was done, e.g., in Ref.~\cite{24} for the 
 $^{181}{\rm{Ta}}(\gamma,3\rm n)^{178}{\rm{Ta}}$  reaction. 

The present paper is concerned with measurements of the total average cross-sections $\langle{\sigma(E_{\rm{\gamma max}})}\rangle$ for photoneutron reactions on the $^{181}$Ta nucleus with emission of up to 8 neutrons at end-point $\gamma$-bremsstrahlung energies $E_{\rm{\gamma max}}$  = 35 to 80 MeV. The obtained results have been compared with our earlier data found for the range $E_{\rm{\gamma max}}$ = 80--95 MeV, and also, with the calculations using the cross-sections $\sigma(E)$ computed with the TALYS1.95 code for different nuclear level density models $LD$~1-6 \cite{9}. 

\section{\label{sec:2} EXPERIMENTAL procedure}

\subsection{\label{sec:2a} Experimental setup}

Experimental tantalum photodisintegration studies have been carried out through measurements of the residual $\gamma$-activity of the irradiated sample, which enabled one to obtain simultaneously the data from different channels of photonuclear reactions.  This technique is well known and has been described in a variety of papers concerned with the investigation of multiparticle photonuclear reactions, e.g., on the nuclei $^{27}$Al \cite{25}, $^{93}$Nb \cite{26,27,28}, $^{181}$Ta \cite{23,24}. 

\begin{figure}[b]
	\resizebox{0.49\textwidth}{!}{%
		\includegraphics{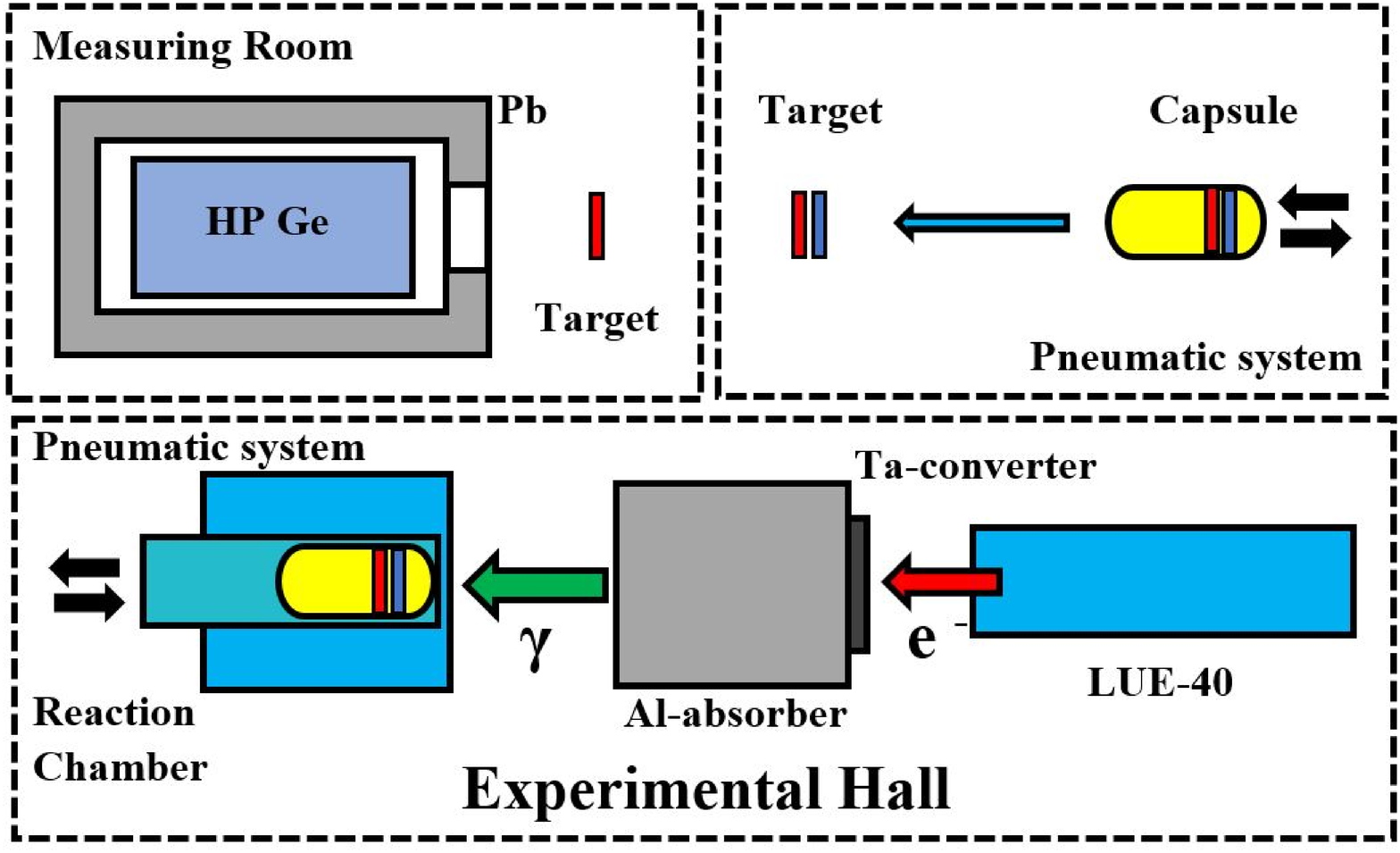}}
	\caption{Schematic block diagram of the experiment. The upper part shows the measuring room, where the exposed target (red color) and the target-monitor (blue color) are extracted from the capsule and are arranged by turn before the HPGe detector for induced $\gamma$-activity measurements. The lower part shows the accelerator LUE-40, the Ta-converter, Al-absorber, exposure chamber.}
	\label{fig1}
\end{figure}

The experimental setup is presented in Fig.~\ref{fig1}. The $\gamma$-ray bremsstrahlung beam was generated by means of the NSC KIPT electron linac LUE-40 RDC “Accelerator” \cite{29,30}. Electrons of the initial energy $E_e$ were incident on the target-converter made from 1.05~mm thick natural tantalum plate, measuring 20 by 20~mm. To remove electrons from the bremsstrahlung flux, a cylindrical aluminum absorber, 100~mm in diameter and 150~mm in length, was used.   
 
The targets of diameter 8~mm, placed in the aluminum capsule, were arranged behind the Al-absorber on the electron beam axis. For transporting the targets to the place of irradiation and back for induced activity registration, the pneumatic tube transfer system was used. On delivery of the irradiated targets to the measuring room, the samples are extracted from the aluminum capsule and are transferred one by one to the detector for the measurements. Taking into account the time of target delivery and extraction from the capsule, the cooling time for the sample under study took no more than 3~minutes.   

The induced $\gamma$-activity of the irradiated targets was registered by the semiconductor HPGe detector Canberra GC-2018 with the resolutions of 0.8 and 1.8~keV (FWHM) for the energies $E_{\gamma}$ = 122 and 1332~keV, respectively. Its efficiency was 20\% relative to the NaI(Tl) detector, 3 inches in diameter and 3 inches in thickness. The absolute registration efficiency of the GC-2018 detector was calibrated with a standard set of $\gamma$-ray radiation sources: $^{22}$Na, $^{60}$Co, $^{133}$Ba, $^{137}$Cs, $^{152}$Eu, $^{241}$Am. 

The bremsstrahlung spectra of electrons were calculated by using the GEANT4.9.2 code \cite{31} with due regard for the real geometry of the experiment, where consideration was given to spatial and energy distributions of the electron beam. The program code GEANT4.9.2 \textit{PhysList G4LowEnergy} allows one to perform calculations taking properly into account all physical processes for the case of an amorphous target. Similarly, GEANT4.9.2 \textit{PhysList QGSP BIC HP} makes it possible to calculate the neutron yield due to photonuclear reactions from targets of different thicknesses and atomic charges. In addition, the bremsstrahlung gamma fluxes were monitored by the yield of the $^{100}\rm{Mo}(\gamma,n)^{99}\rm{Mo}$ reaction. For this purpose, the natural molybdenum target-witness, placed close by the target under study, was simultaneously exposed to radiation. 

In the experiment, natural tantalum/molybdenum samples were exposed to radiation at end-point bremsstrahlung energies $E_{\rm \gamma max}$ ranging from 35 to 80~MeV with an energy step of $\sim$5~MeV.  The masses of tantalum and molybdenum targets were, respectively, $\sim$43~mg and $\sim$60~mg.  The time of irradiation $t_{\rm{irr}}$ and the time of residual $\gamma$-activity spectrum measurement $t_{\rm{meas}}$ were both 30~min. 

\begin{figure*}[]
	\begin{minipage}[h]{0.95\linewidth}
		\center{\includegraphics[width=1\linewidth]{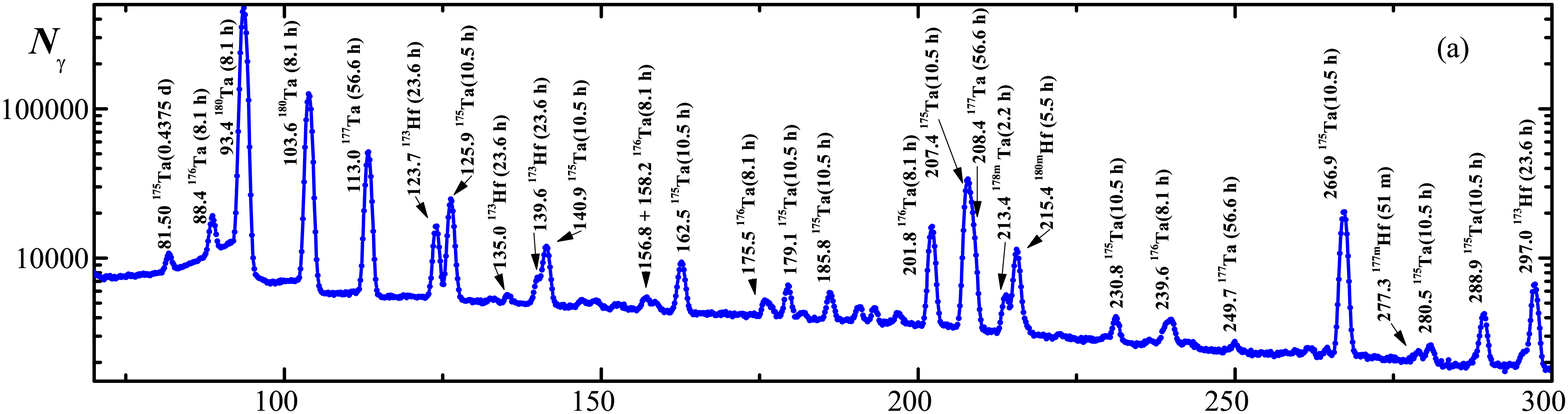}} \\
	\end{minipage}
	\vfill
	\begin{minipage}[h]{0.95\linewidth}
		\center{\includegraphics[width=1\linewidth]{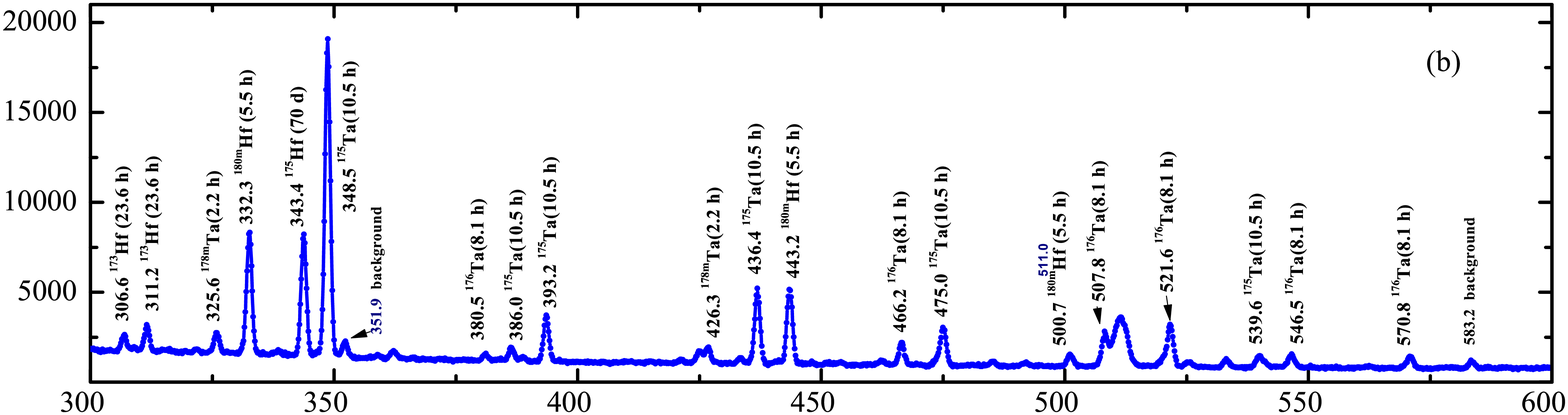}} \\
	\end{minipage}
	\begin{minipage}[h]{0.95\linewidth}
	\center{\includegraphics[width=1\linewidth]{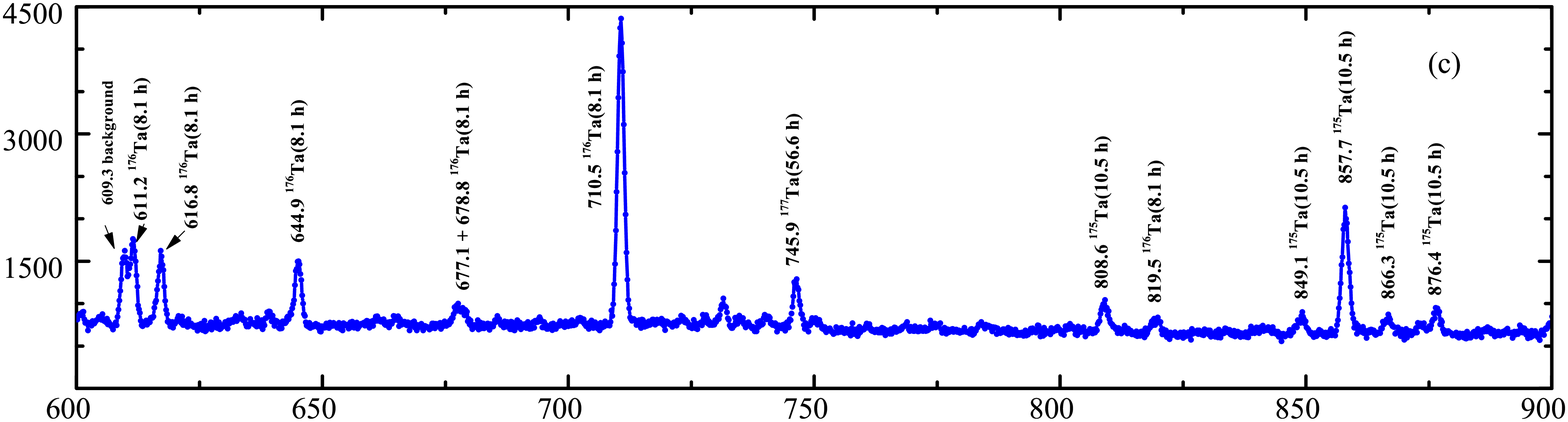}} \\
	\end{minipage}

	\begin{minipage}[h]{0.95\linewidth}
	\center{\includegraphics[width=1\linewidth]{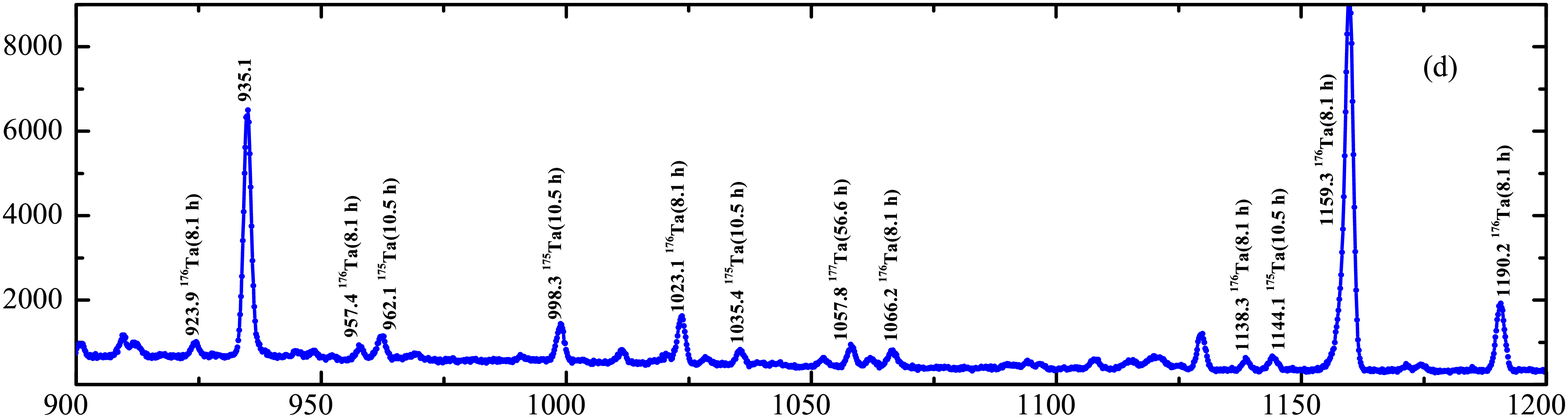}} \\
	\end{minipage}
	\begin{minipage}[h]{0.95\linewidth}
	\center{\includegraphics[width=1\linewidth]{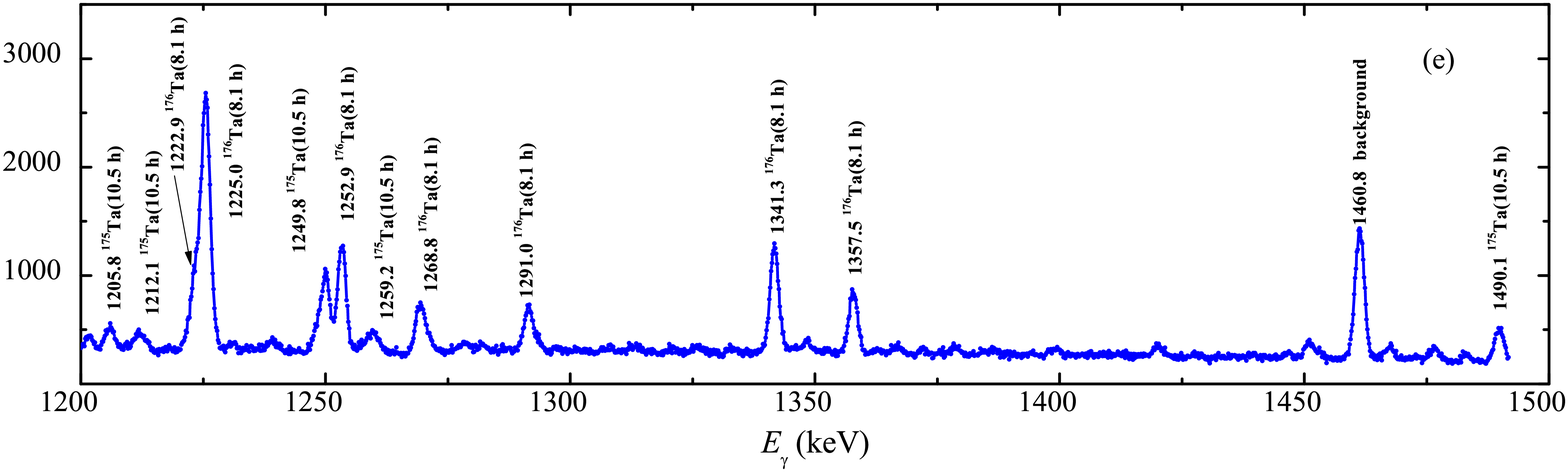}} \\
\end{minipage}
	\caption{$\gamma$-ray spectrum from the  43.7~mg $^{181}\rm Ta$ target after exposure for 30 min to  the bremsstrahlung $\gamma$-flux at  $E_{\rm{\gamma max}}$ = 80.7~MeV.}
	\label{fig2}
\end{figure*}

Figure~\ref{fig2} shows the long-term measurement  gamma spectrum from reaction products of the tantalum target in the $E_{\gamma}$ range from 40 to 1500~keV. 

Table~\ref{tab1} lists the parameter values of the reaction $^{181}\rm{Ta}(\gamma,\textit{x}n; \textit{x} \leq 8)^{181-\textit{x}}\rm{Ta}$ and 
 the monitoring reaction $^{100}\rm{Mo}(\gamma,n)^{99}\rm{Mo}$ according to the data from \cite{32}: $E_{\rm{th}}$ – reaction
  thresholds; $J$, $\pi$, $T_{1/2}$ denote the spin, parity and half-life period of the nuclei-products, respectively;
 $E_{\gamma}$ are the energies of the $\gamma$-lines under study and their intensities $I_{\gamma}$. 

  \begin{table*}[]
	\caption{\label{tab1} Spectroscopic data \cite{32} on the nuclei-products from the reactions $^{181}\rm{Ta}(\gamma,\textit{x}n; \textit{x} = 1$--$8)^{181-\textit{x}}\rm{Ta}$ and the monitoring reaction $^{100}\rm{Mo}(\gamma,n)^{99}\rm{Mo}$}
	\centering
	\begin{ruledtabular}
		\begin{tabular}{cccccc}
			\begin{tabular}{c} Nuclear \\  reaction \end{tabular} & $E_{\rm{th}}$,~MeV & \begin{tabular}{c} $J^\pi$ of  \\ nucleus-product
			\end{tabular} & $T_{1/2}$ & $E_{\gamma}$,~keV & $I_{\gamma}$, \% \\ 	\hline
			&&&&&\\ 	   
			$^{181}\rm{Ta}(\gamma,n)^{180g}\rm{Ta}$ & 7.58 & $1^+$ & 8.152$\pm$0.006 h & 103.557$\pm$0.007 & 0.81$\pm$0.16 \\
			$^{181}\rm{Ta}(\gamma,n)^{180m}\rm{Ta}$ & 7.65 & $9^-$ & $>1.2 \cdot 10^{15}$ y &   &   \\ 
			$^{181}\rm{Ta}(\gamma,2n)^{179}\rm{Ta}$ & 14.22 & $7/2^+$ & 1.82$\pm$0.03 y & 54.611  & 13.6$\pm$0.4   \\
			&&&&  55.790 & 23.7$\pm$0.7 \\
			$^{181}\rm{Ta}(\gamma,3n)^{178g}\rm{Ta}$ & 22.05 & $1^+$ & 9.31$\pm$0.03 min & 1350.68$\pm$0.03 & 1.18$\pm$0.03 \\
			$^{181}\rm{Ta}(\gamma,3n)^{178m}\rm{Ta}$ & 22.35 & $(7)^-$ & 2.36$\pm$0.08 h & 426.383$\pm$0.006 & 97.0$\pm$1.3 \\
			$^{181}\rm{Ta}(\gamma,4n)^{177}\rm{Ta}$ & 29.01 & $ 7/2^+$ & 56.56$\pm$0.06 h & 112.9498$\pm$0.0005 & 7.2$\pm$0.8 \\
			$^{181}\rm{Ta}(\gamma,5n)^{176}\rm{Ta}$ & 37.44 & $ (1)^-$ & 8.09$\pm$0.05 h & 1159.28$\pm$0.09 & 25.00$\pm$0.15\footnote{Error of intensity $I_{\gamma}$ for the 1159.28~keV and 172.2~keV $\gamma$-lines were determined as half-value spreads according to the databases of \cite{32} and \cite{33}.} \\
			$^{181}\rm{Ta}(\gamma,6n)^{175}\rm{Ta}$ & 44.46 & $ 7/2^+$ & 10.5$\pm$0.2 h & 348.5$\pm$0.5 & 12.0$\pm$0.6 \\
			$^{181}\rm{Ta}(\gamma,7n)^{174}\rm{Ta}$ & 53.21 & $ 3^+$ & 1.05$\pm$0.03 h & 1205.92$\pm$0.04 & 4.9$\pm$0.4 \\
			$^{181}\rm{Ta}(\gamma,8n)^{173}\rm{Ta}$ & 60.63 & $ 5/2^-$ & 3.14$\pm$0.13 h & 172.2$\pm$0.1 & $18.00\pm0.15^{\text{a}}$ \\
			$^{100}\rm{Mo}(\gamma,n)^{99}\rm{Mo}$   & 8.29   & $1/2^+$  & 65.94$\pm$0.01 h & 739.50$\pm$0.02 & 12.13$\pm$0.12  \\ 
		\end{tabular}
	\end{ruledtabular}	        
\end{table*}
\vspace{2ex}

\subsection{\label{sec:2b} Calculation formulas for the flux-average cross-sections}

The cross-sections $\langle{\sigma(E_{\rm{\gamma max}})}\rangle$ averaged over the bremsstrahlung $\gamma$-flux $W(E,E_{\rm{\gamma max}})$ from the threshold $E_{\rm{th}}$ of the reaction understudy to the end-point energy of the spectrum $E_{\rm{\gamma max}}$ were calculated with the use of the theoretical cross-section $\sigma(E)$ values computed with the TALYS1.95 code \cite{9}. The theoretical bremsstrahlung flux-averaged cross-section $\langle{\sigma(E_{\rm{\gamma max}})}\rangle$ in a given energy interval and the experimental ones were calculated by the formulas \ref{form1} and \ref{form2}, respectively:

\begin{widetext}
 \begin{equation}\label{form1}
\langle{\sigma(E_{\rm{\gamma max}})}\rangle =   {\rm{\Phi}}^{-1}(E_{\rm{\gamma max}}) \int\limits_{E_{\rm{th}}}^{E_{\rm{\gamma max}}}\sigma(E)\cdot W(E,E_{\rm{\gamma max}})dE.
\end{equation}
  	\begin{equation}
\langle{\sigma(E_{\rm{\gamma max}})}\rangle = \\
\frac{\lambda \triangle A  }{N_n I_{\gamma} \ \varepsilon\ {\rm{\Phi}}(E_{\rm{\gamma max}}) (1-\exp(-\lambda t_{\rm{irr}}))\exp(-\lambda t_{\rm{cool}})(1-\exp(-\lambda t_{\rm{meas}}))},
\label{form2}
 	\end{equation}
 \end{widetext}
where $\triangle A$ is the number of counts of ${\gamma}$-quanta in the full absorption peak, $\lambda$ is the decay constant \mbox{($\rm{ln}2/\textit{T}_{1/2}$)}, $N_n$ is the number of target atoms, $I_{\gamma}$ is the absolute intensity of the analyzed ${\gamma}$-quanta, $\varepsilon$ is the absolute detection efficiency for the analyzed ${\gamma}$-quanta energy, ${\rm{\Phi}}(E_{\rm{\gamma max}}) = {\int\limits_{E_{\rm{th}}}^{E_{\rm{\gamma max}}}W(E,E_{\rm{\gamma max}})dE}$ is the integrated bremsstrahlung flux $W(E,E_{\rm{\gamma max}})$ in the energy range from the reaction threshold $E_{\rm{th}}$ of the corresponding reaction up to the maximum energy of ${\gamma}$-quanta $E_{\rm{\gamma max}}$; $t_{\rm{irr}}$, $t_{\rm{cool}}$ and $t_{\rm{meas}}$ are the irradiation time, cooling time and measurement time, respectively. A more detailed description of all the calculation procedures necessary for the determination of $\langle{\sigma(E_{\rm{\gamma max}})}\rangle$ can be found in \cite{27,34}. 

If the nucleus-product has the isomeric state, the value of the total average cross-section for the reaction under study $\langle{\sigma(E_{\rm{\gamma max}})}\rangle_{\rm{tot}}$ (hereinafter referred to as $\langle{\sigma(E_{\rm{\gamma max}})}\rangle$) is calculated as the sum of $\langle{\sigma(E_{\rm{\gamma max}})}\rangle_{\rm{g}}$ and $\langle{\sigma(E_{\rm{\gamma max}})}\rangle_{\rm{m}}$, being, respectively, the average cross-sections for ground-state and isomeric-state.

In our calculations of average reaction cross-sections, all radioactive isotopes were assumed to be produced as a result of photonuclear reactions on $^{181}\rm{Ta}$, considering that the $^{180\rm{m}}\rm{Ta}$ content of the natural tantalum mixture is negligibly small (0.012\%). The self-absorption of reaction-product $\gamma$-rays in the target was computed in the GEANT4.9.2 code. 

\subsection{\label{sec:2c} Bremsstrahlung $\gamma$-flux monitoring}

\begin{figure}[h]
	\resizebox{0.49\textwidth}{!}{%
		\includegraphics{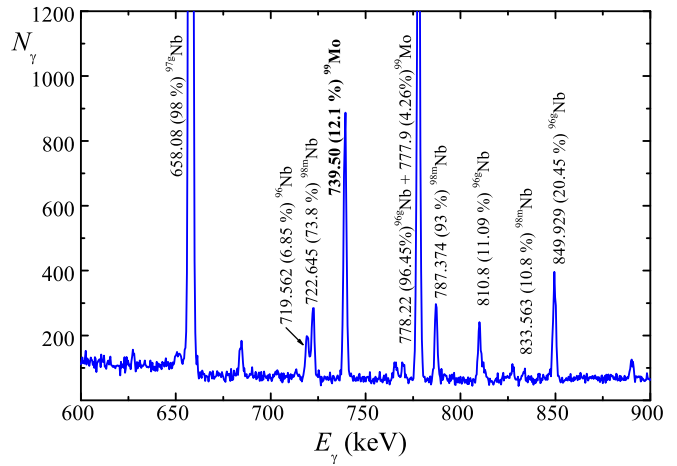}}
	\caption{Fragment of the $\gamma$-ray spectrum from the 56.7~mg Mo-target after its exposure to the bremsstrahlung $\gamma$-flux at $E_{\gamma \rm max}$ = 80.7~MeV for $t_{\rm irr}$  = 30~min. 600 $\leq E_\gamma \leq$ 900~keV.}
	\label{fig3}
\end{figure}

The bremsstrahlung gamma flux monitoring against the $^{100}\rm{Mo}(\gamma,n)^{99}\rm{Mo}$ reaction yield was performed by comparing the experimentally obtained average cross-section values with the computation data. To determine the experimental $\langle{\sigma(E_{\rm{\gamma max}})}\rangle_{\rm{exp}}$ values by Eq.~\ref{form2}, we have used the $\triangle A$ activity value for the $E_\gamma$ = 739.50~keV $\gamma$-line and the absolute intensity $I_{\gamma}$ = 12.13\% (see Table~\ref{tab1}, example spectrum in Fig.~\ref{fig3}). The theoretical values of the average cross-section $\langle{\sigma(E_{\rm{\gamma max}})}\rangle_{\rm{th}}$ were calculated by Eq.~\ref{form1} using the cross-sections $\sigma(E)$ with the TALYS1.95 code with the default options. The normalization (monitoring) factor $k_{\rm{monitor}}$, derived from the ratios of $\langle{\sigma(E_{\rm{\gamma max}})}\rangle_{\rm{th}}$ to $\langle{\sigma(E_{\rm{\gamma max}})}\rangle_{\rm{exp}}$, represent the deviation of the GEANT4.9.2-computed bremsstrahlung $\gamma$-flux from the real flux falling on the target. The determined $k_{\rm{monitor}}$ values were used for normalizing the cross-sections for photonuclear reactions on the $^{181}\rm{Ta}$ nucleus. The monitoring procedure has been detailed in \cite{26,27}.

The Ta-converter and Al-absorber, used in the experiment, generate the neutrons that can cause the reaction $^{100}\rm{Mo}(n,2n)^{99}\rm{Mo}$. Calculations were made of the neutron energy spectrum and the fraction of neutrons of energies above the threshold of this reaction, similarly to \cite{35}. The contribution of the $^{100}\rm{Mo}(n,2n)^{99}\rm{Mo}$ reaction to the induced activity of the $^{99}\rm{Mo}$ nucleus has been estimated and is negligible as compared to the contribution of $^{100}\rm{Mo}(\gamma,n)^{99}\rm{Mo}$. The contribution of the reaction $^{100}\rm{Mo}(\gamma,p)^{99}\rm{Nb}$, $^{99}\rm{Nb} \xrightarrow{\beta^-}$$^{99}\rm{Mo}$ is also negligible.

\subsection{\label{sec:2d} Experimental accuracy of the average cross-sections $\langle{\sigma(E_{\rm{\gamma max}})}\rangle$, $\langle{\sigma(E_{\rm{\gamma max}})}\rangle_{\rm{g}}$ and $\langle{\sigma(E_{\rm{\gamma max}})}\rangle_{\rm{m}}$}

The uncertainty in measurements of experimental values of the average cross-sections  $\langle{\sigma(E_{\rm{\gamma max}})}\rangle$, $\langle{\sigma(E_{\rm{\gamma max}})}\rangle_{\rm{g}}$ and $\langle{\sigma(E_{\rm{\gamma max}})}\rangle_{\rm{m}}$  was determined as a quadratic sum of statistical and systematical errors. The statistical error in the observed $\gamma$-activity is mainly due to statistics calculation in the total absorption peak of the corresponding $\gamma$-line, which varies within 1 to 10\%. This error varies depending on the $\gamma$-line intensity and the background conditions of spectrum measurements. The intensity of the line depends on the detection efficiency, the half-life period, and the absolute intensity $I_\gamma$. The background is generally governed by the contribution of the Compton scattering of quanta. 

The systematical errors are due to the following uncertainties: 
\begin{enumerate} 
	\item	time of exposure and the electron current $\sim$0.5\%; 
	\item  $\gamma$-ray registration efficiency of the detector, $\sim$2--3\%, which is generally associated with the gamma radiation source error. The error is greater at $\gamma$-quantum energies $E_\gamma$ = 50--200~keV, this being due to a small $\gamma$-line quantity in this energy range and the intricate shape of the efficiency curve;
	\item	the half-life time $T_{1/2}$ of the reaction products and the absolute intensity of the analyzed $\gamma$-quanta $I_{\gamma}$ = 1.3--20\%, as is noted in Table~\ref{tab1} according to the data from~\cite{32}. For the case of $^{181}\rm{Ta}(\gamma,5n)^{176}Ta$ and $^{181}\rm{Ta}(\gamma,8n)^{173}Ta$ reactions, the errors of $\gamma$-line intensities were determined as half-value spreads from the databases of \cite{32} and \cite{33}. The errors amounted to 0.6\% and 1.4\% for 1159.28~keV ($^{176}\rm{Ta}$) and 172.2~keV ($^{176}\rm{Ta}$), respectively;
	\item	normalization of the experimental data to the yield of the monitoring reaction   $^{100}\rm{Mo}(\gamma,n)^{99}\rm{Mo}$  made up to 4\%. It should be noted that the systematic error in yield monitoring of the $^{100}\rm{Mo}(\gamma,n)^{99}\rm{Mo}$ reaction stems from three unavoidable errors, each running to $\sim$1\%. These are the unidentified isotopic composition of natural molybdenum, the uncertainty in the $\gamma$-line intensity used, $I_{\gamma}$, and the statistical error in the determination of the area under the normalizing $\gamma$-line peak. In our calculations, we have used the percentage value of $^{100}\rm{Mo}$ isotope abundance equal to 9.63\% (see Ref.~\cite{31});
	\item	the error of $\gamma$-flux computation with the GEANT4.9.2 code ranged within $\sim$1--1.5\%.
\end{enumerate}

Thus, the statistical and systematical errors are the variables, which differ for different $^{181}\rm{Ta}(\gamma,\textit{x}n; \textit{x} \leq 8)^{181-\textit{x}}\rm{Ta}$ reactions. The total uncertainty of the experimental data is given in Fig.~\ref{fig5} and Tables~\ref{tab2},\ref{tab3},\ref{tab4}.

\section{\label{sec:3} $^{181}\rm{Ta}(\gamma,\textit{x}n; \textit{x} \leq 8)^{181-\textit{x}}\rm{Ta}$ REACTION CROSS-SECTIONS RESULTING FROM THE TALYS1.95 CODE COMPUTATIONS}

The theoretical values of total and partial cross-sections $\sigma(E)$ for the $^{181}\rm{Ta}(\gamma,\textit{x}n)^{181-\textit{x}}\rm{Ta}$ reactions with emission of up to 8 neutrons  were taken for the monochromatic photons from the computations by the TALYS1.95 code \cite{9} installed on the Linux Ubuntu-20.04. The computations were performed for different level density models \textit{LD} 1--6. There are 3 phenomenological level density models and 3 options for microscopic level densities:

$LD 1$: Constant temperature + Fermi gas model. In this model introduced by Gilbert and Cameron, the excitation energy range is divided into a low energy part from $E_0$ up to a matching energy $E_{\rm{M}}$, where the so-called constant temperature law applies and a high energy part above, where the Fermi gas model applies. 

$LD 2$: Back-shifted Fermi gas model. In the Back-shifted Fermi gas model, the pairing energy is treated as an adjustable parameter and the Fermi gas expression is used down to $E_0$.

$LD 3$: Generalized superfluid model (GSM). The model takes superconductive pairing correlations into account according to the Bardeen-Cooper-Schrieffer theory. The phenomenological version of the model is characterized by a phase transition from a superfluid behavior at low energy, where pairing correlations strongly influence the level density, to a high energy region which is described by the Fermi gas model. The GSM thus resembles the constant temperature model to the extent that it distinguishes between low energy and a high energy region, although for the GSM this distinction follows naturally from the theory and does not depend on specific discrete levels that determine matching energy. Instead, the model automatically provides a constant temperature-like behavior at low energies.

$LD 4$: Microscopic level densities (Skyrme force) from Goriely’s tables. Using this model allows reading tables of microscopic level densities from RIPL database. These tables were computed by S. Gorielyon based on Hartree-Fock calculations for excitation energies up to 150~MeV and for spin values up to $I$ = 30. 

$LD 5$: Microscopic level densities (Skyrme force) from Hilaire’s combinatorial tables. The combinatorial model includes a detailed microscopic calculation of the intrinsic state density and collective enhancement. The only phenomenological aspect of the model is a simple damping function for the transition from spherical to deformed.  

$LD 6$: Microscopic level densities (temperature-dependent HFB, Gogny force) from Hilaire’s combinatorial tables.

\begin{figure*}[]
	\begin{minipage}[h]{0.49\linewidth}
		\center{\includegraphics[width=1\linewidth]{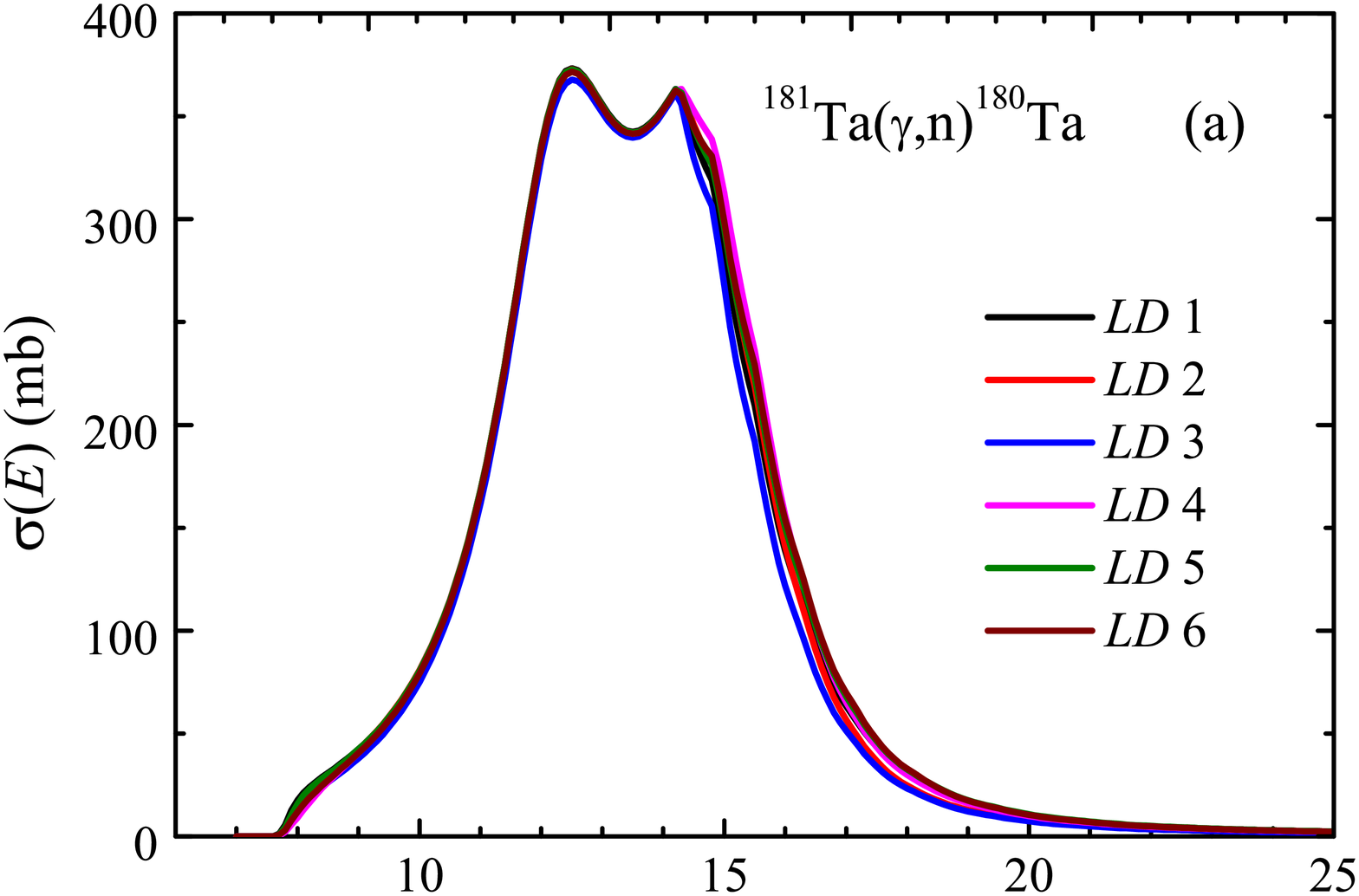}} \\
	\end{minipage}
	\hfill
	\begin{minipage}[h]{0.49\linewidth}
		\center{\includegraphics[width=1\linewidth]{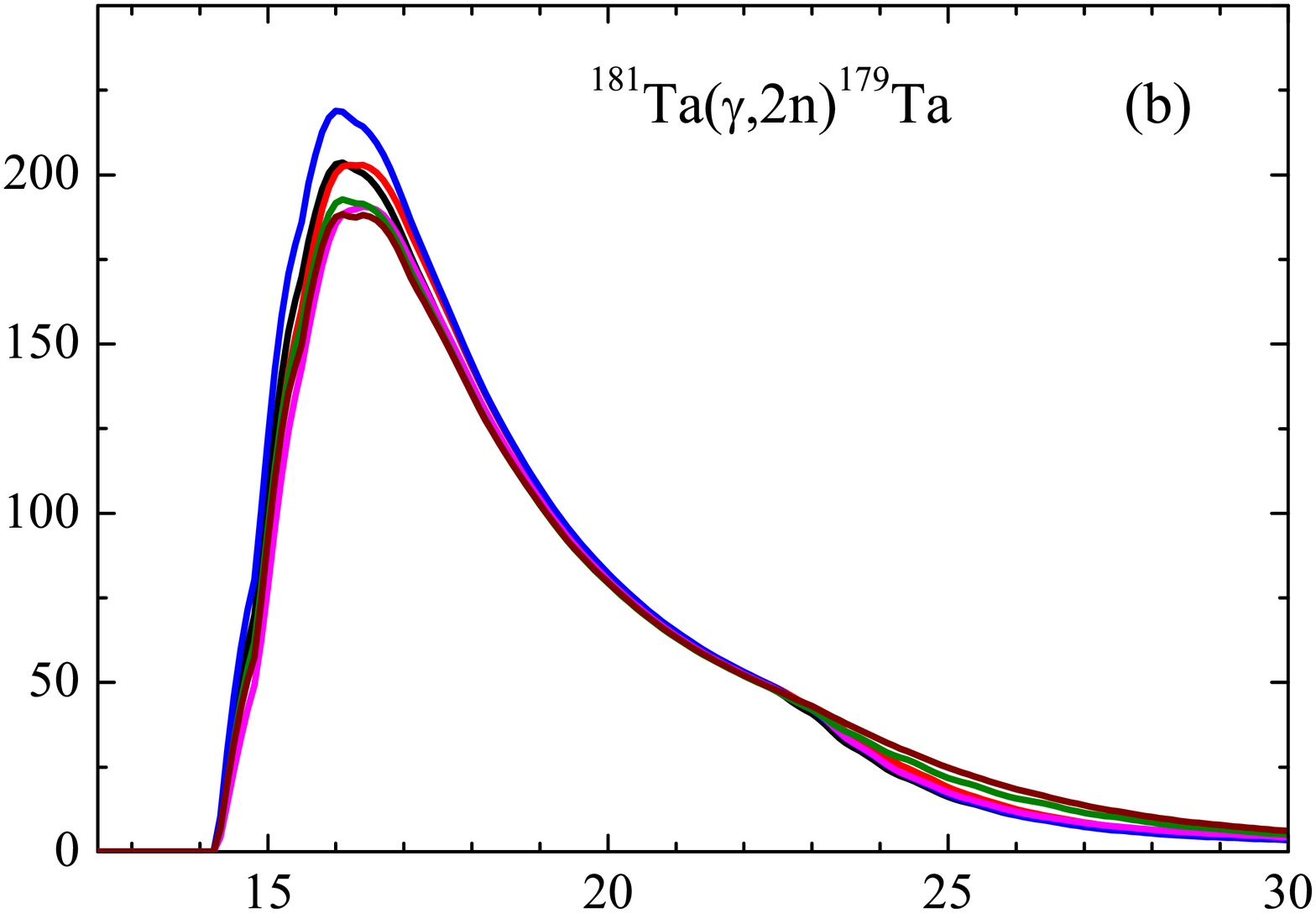}} \\
	\end{minipage}
	\vfill
	\begin{minipage}[h]{0.49\linewidth}
		\center{\includegraphics[width=1\linewidth]{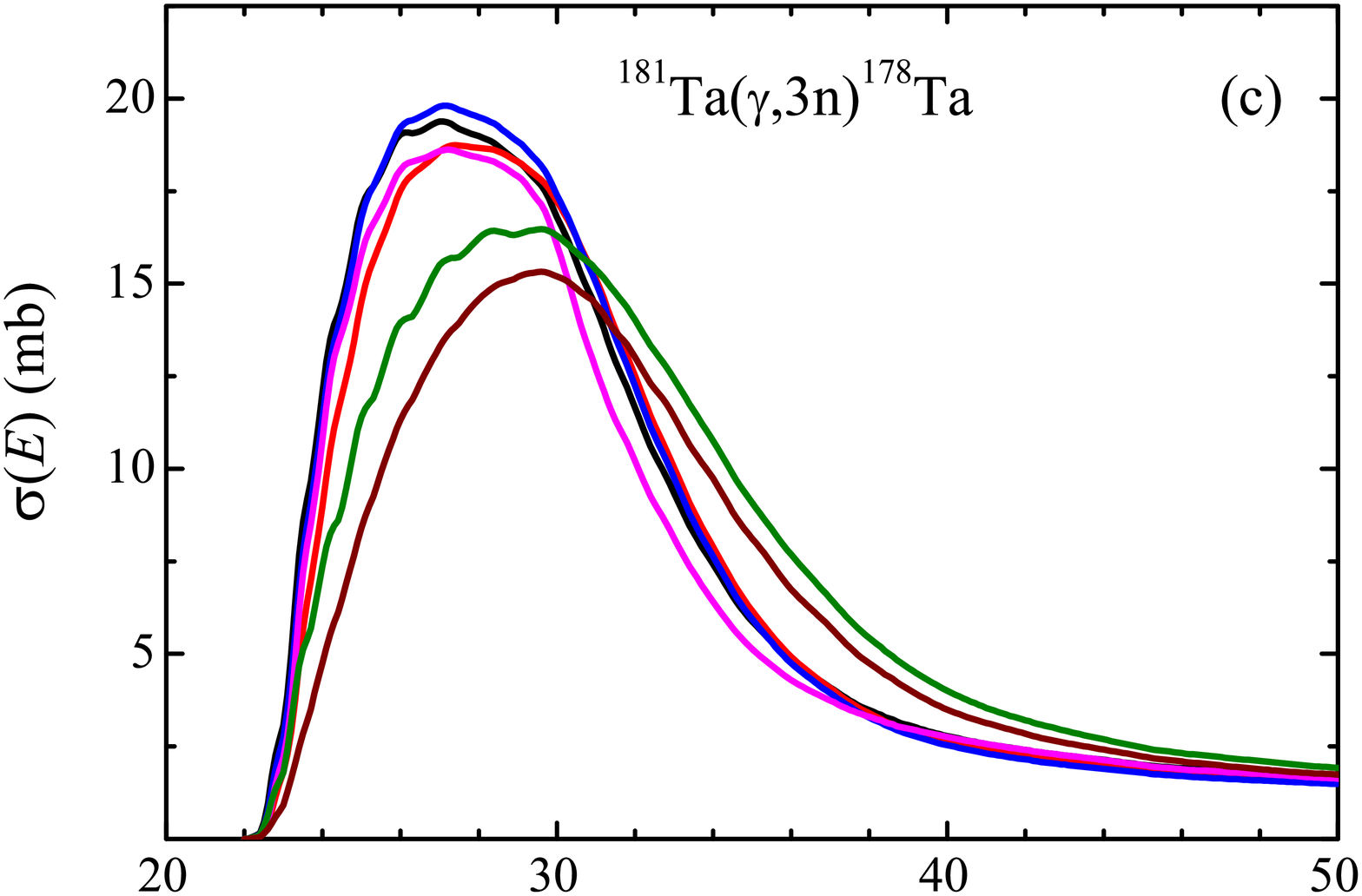}}\\
	\end{minipage}
	\hfill
	\begin{minipage}[h]{0.49\linewidth}
		\center{\includegraphics[width=1\linewidth]{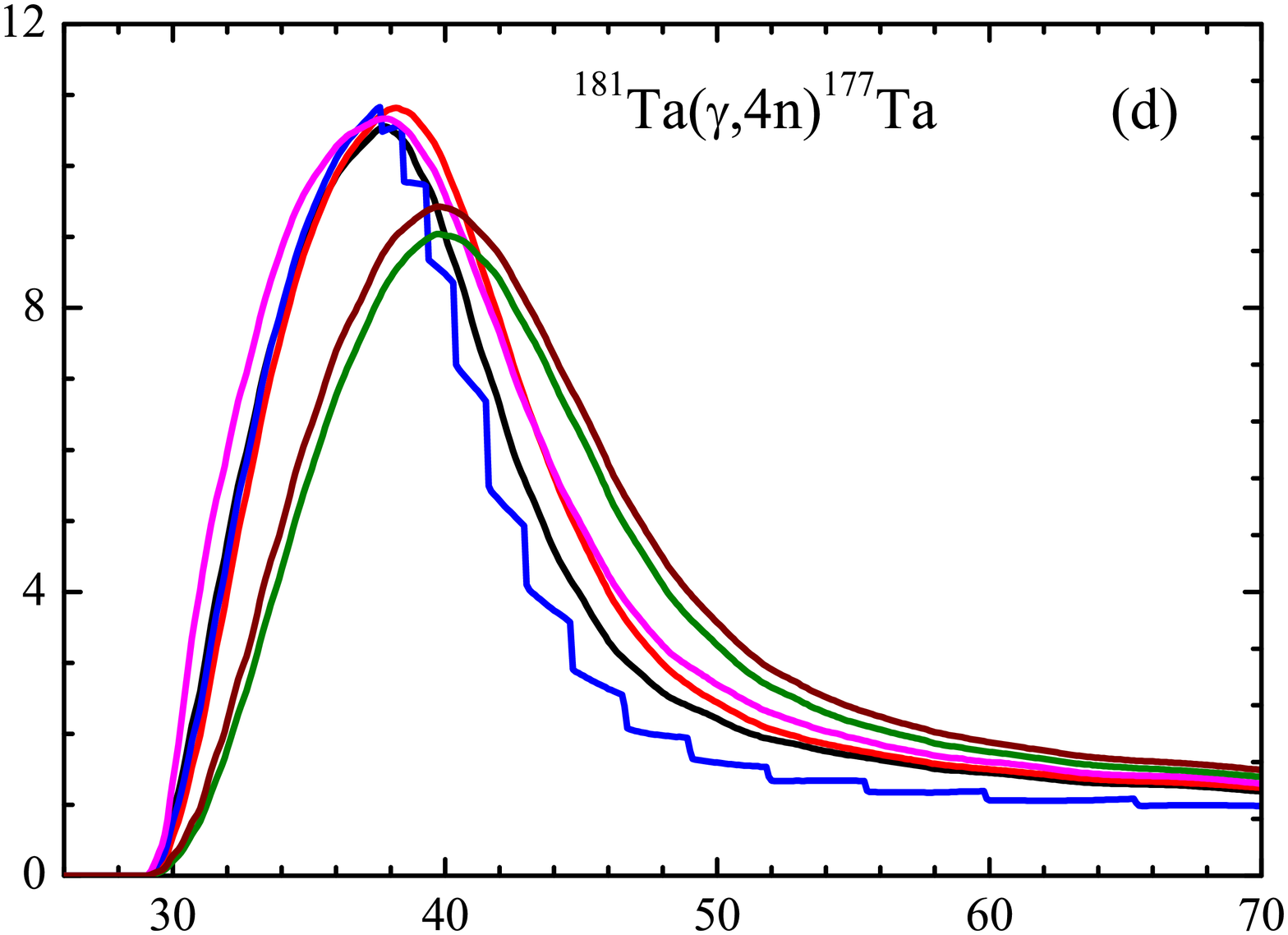}}\\
	\end{minipage}
	\begin{minipage}[h]{0.49\linewidth}
	\center{\includegraphics[width=1\linewidth]{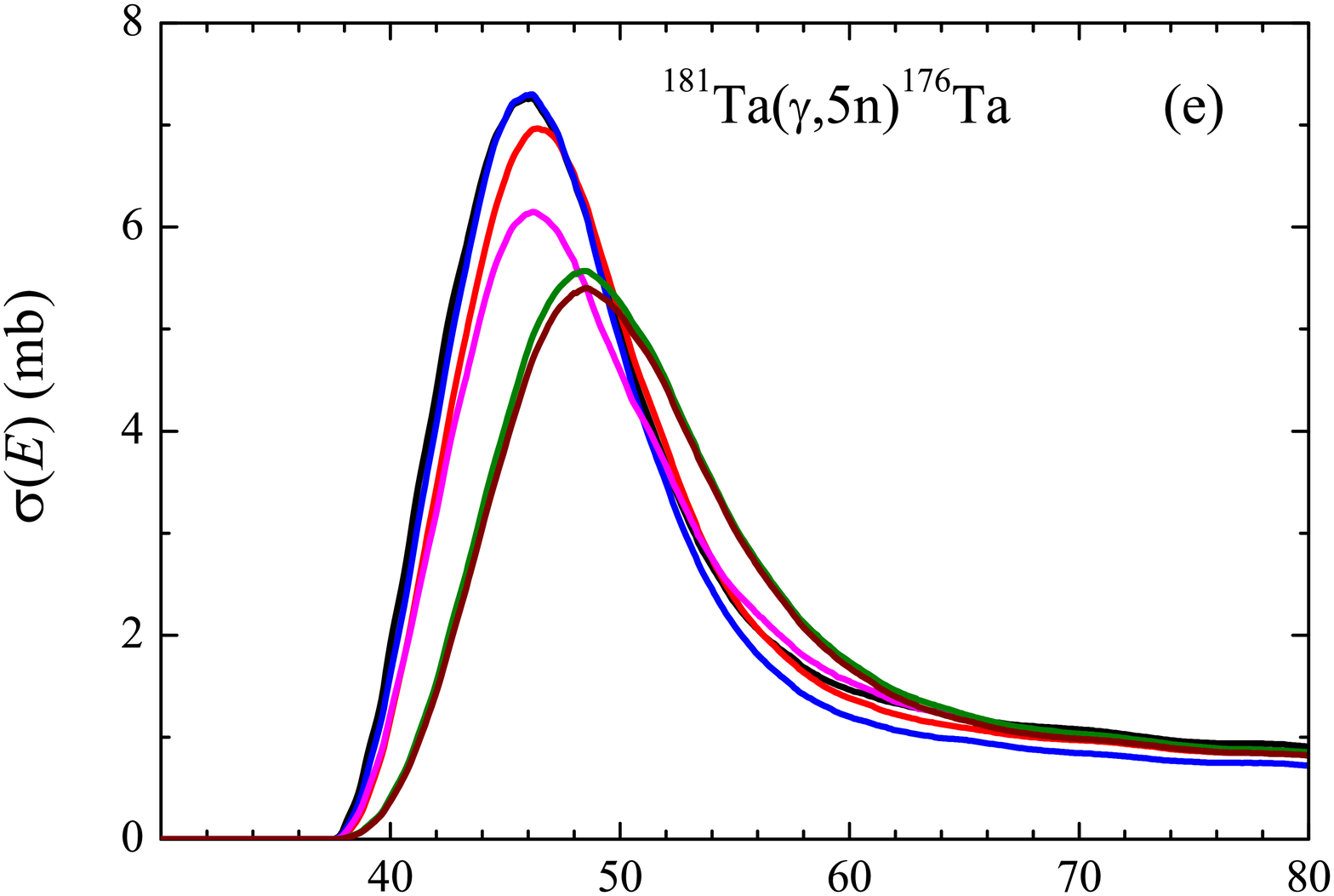}} \\
	\end{minipage}
	\hfill
	\begin{minipage}[h]{0.49\linewidth}
		\center{\includegraphics[width=1\linewidth]{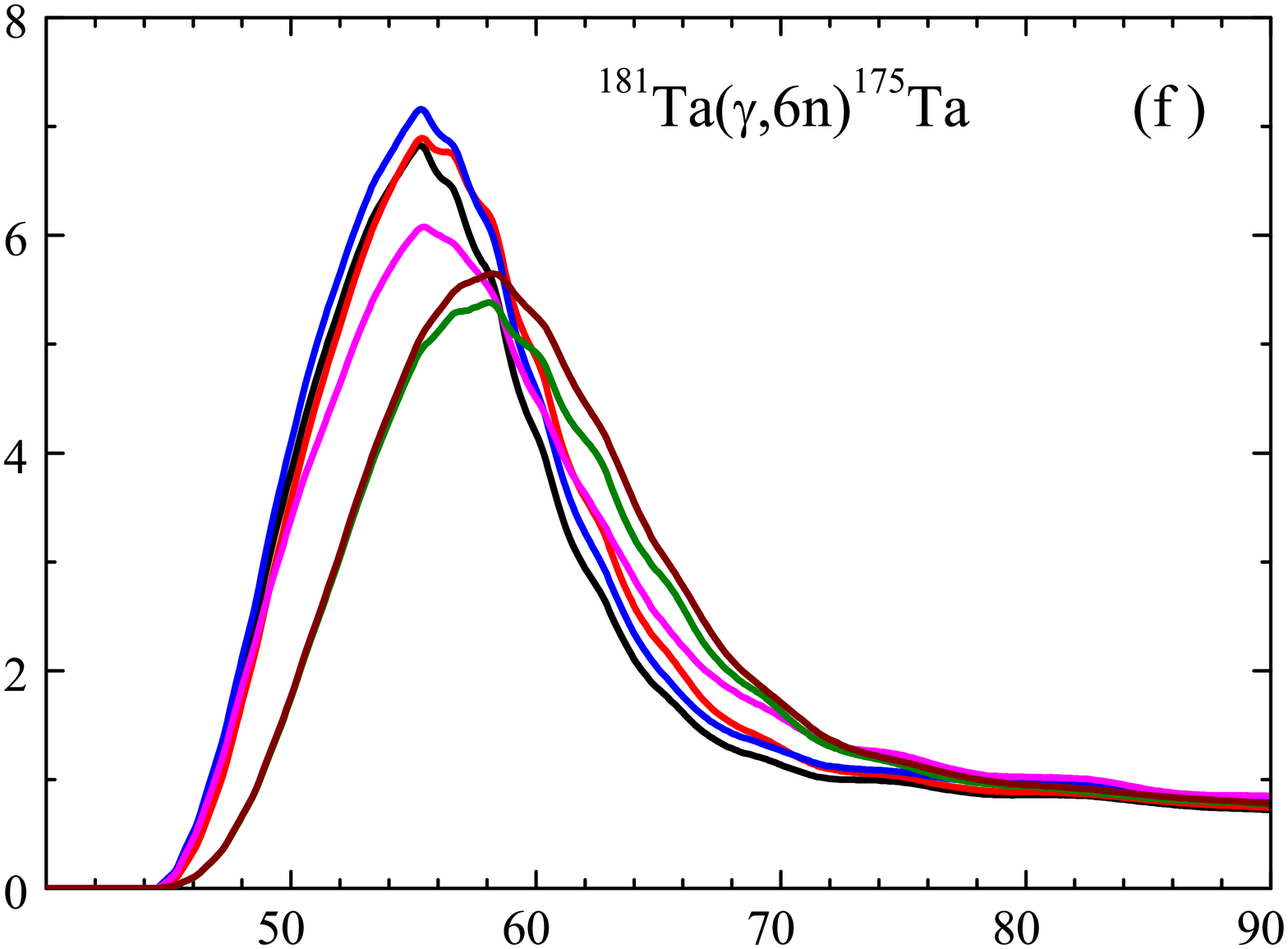}} \\
	\end{minipage}
	\vfill
	\begin{minipage}[h]{0.49\linewidth}
		\center{\includegraphics[width=1\linewidth]{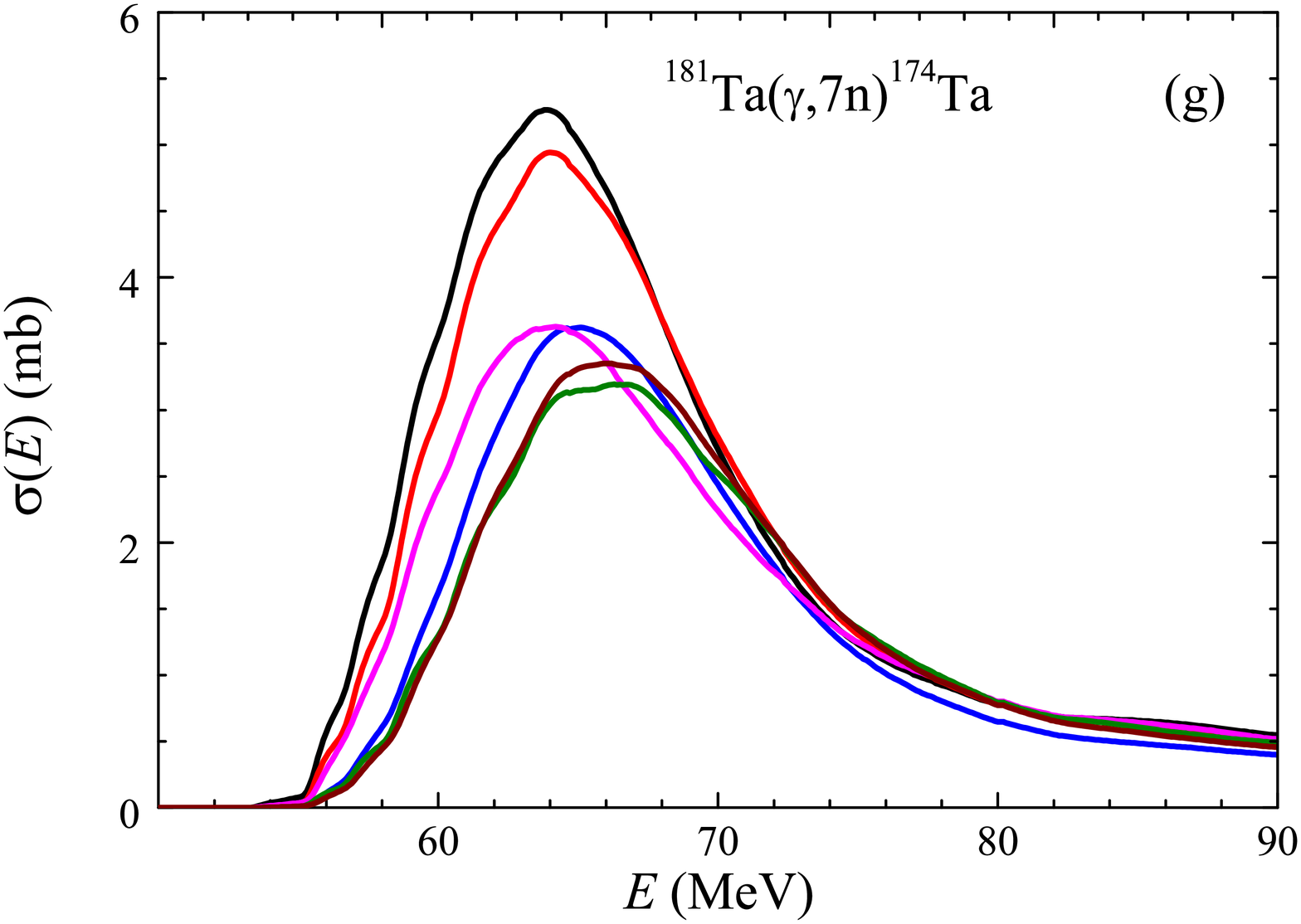}}\\
	\end{minipage}
	\hfill
	\begin{minipage}[h]{0.49\linewidth}
		\center{\includegraphics[width=1\linewidth]{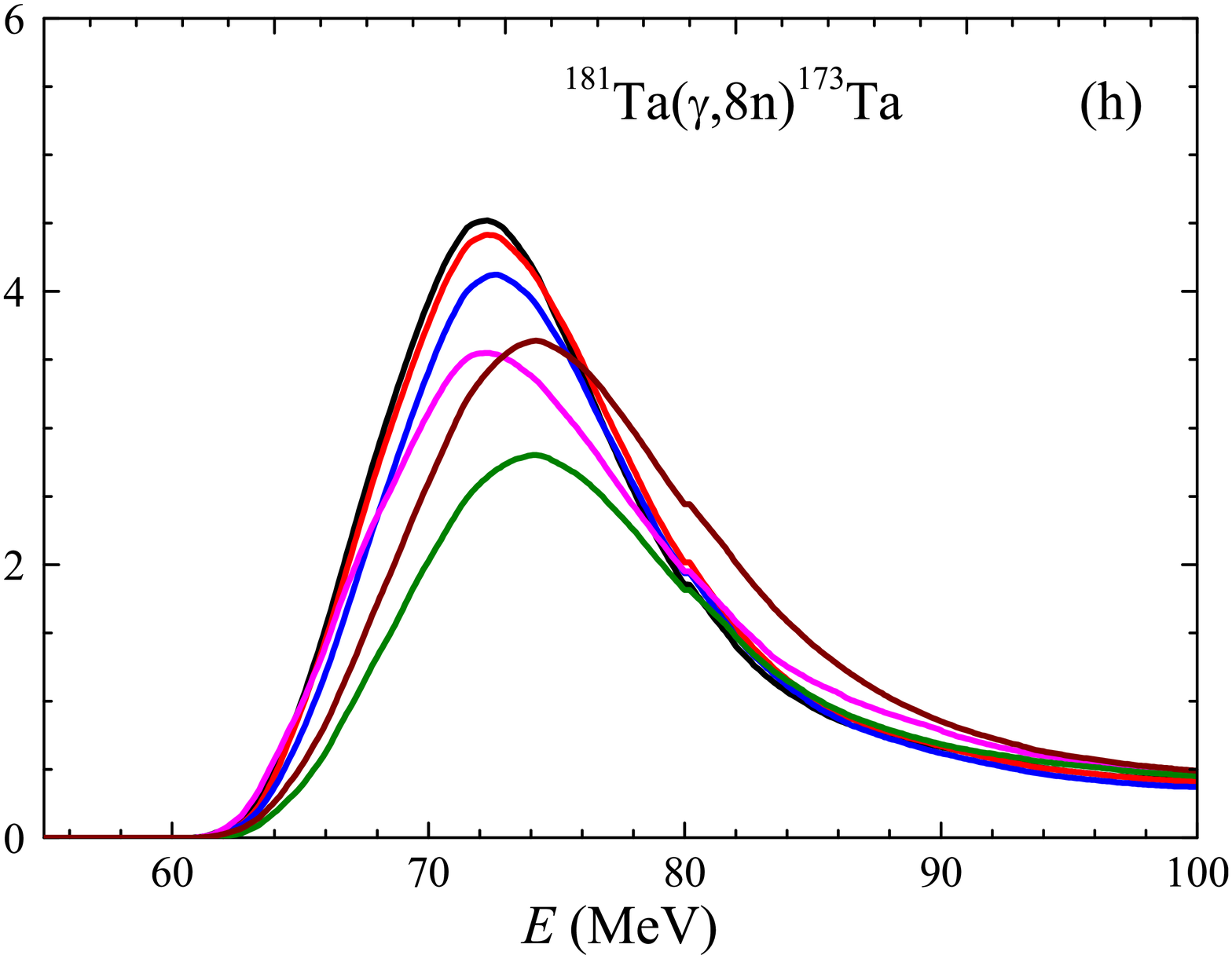}}\\
	\end{minipage}
	\caption{Total cross-sections $\sigma(E)$ for the  $^{181}\rm{Ta}(\gamma,\textit{x}n; \textit{x} \leq 8)^{181-\textit{x}}\rm{Ta}$ reactions from  TALYS1.95 computations for different level density models \textit{LD} 1-6.}
	\label{fig4}
\end{figure*}

The TALYS1.95 computation data on the total cross-sections for the $^{181}\rm{Ta}(\gamma,\textit{x}n; \textit{x} \leq 8)^{181-\textit{x}}\rm{Ta}$ reactions are presented in Fig.~\ref{fig4}. The data were obtained for different level density models. It is obvious from the figures that for the $(\gamma,\rm{n})$ and $(\gamma,\rm{2n})$ reactions the variants of calculations for \textit{LD} 1-6 are closely located. With the increase in the number of neutrons in the outlet channel, the behavior of the cross-sections with energy for the \textit{LD}5 and \textit{LD}6 models is different from that for the other models, viz., the maxima of cross-sections are somewhat lower, while their positions are shifted towards higher energy.

\section{\label{sec:4} RESULTS AND DISCUSSION}

In the present work, we obtained the experimental values of the total average cross-sections $\langle{\sigma(E_{\rm{\gamma max}})}\rangle$ for the $^{181}\rm{Ta}(\gamma,\textit{x}n; \textit{x} \leq 8)^{181-\textit{x}}\rm{Ta}$ reaction in the range of end-point bremsstrahlung $\gamma$-quantum energies $E_{\rm{\gamma max}}$ = 35--80~MeV. The found cross-sections are compared with the theoretical values of the average cross-sections calculated by Eq.~\ref{form1} using bremsstrahlung fluxes corresponding to real experimental conditions and cross-sections from TALYS1.95 for level density models \textit{LD} 1-6. Thus, each cross-section $\langle{\sigma(E_{\rm{\gamma max}})}\rangle$ of the $^{181}\rm{Ta}(\gamma,\textit{x}n; \textit{x} \leq 8)^{181-\textit{x}}\rm{Ta}$ reactions corresponds to six variants of calculation.

\begin{figure*}[]
	\begin{minipage}[h]{0.49\linewidth}
		\center{\includegraphics[width=1\linewidth]{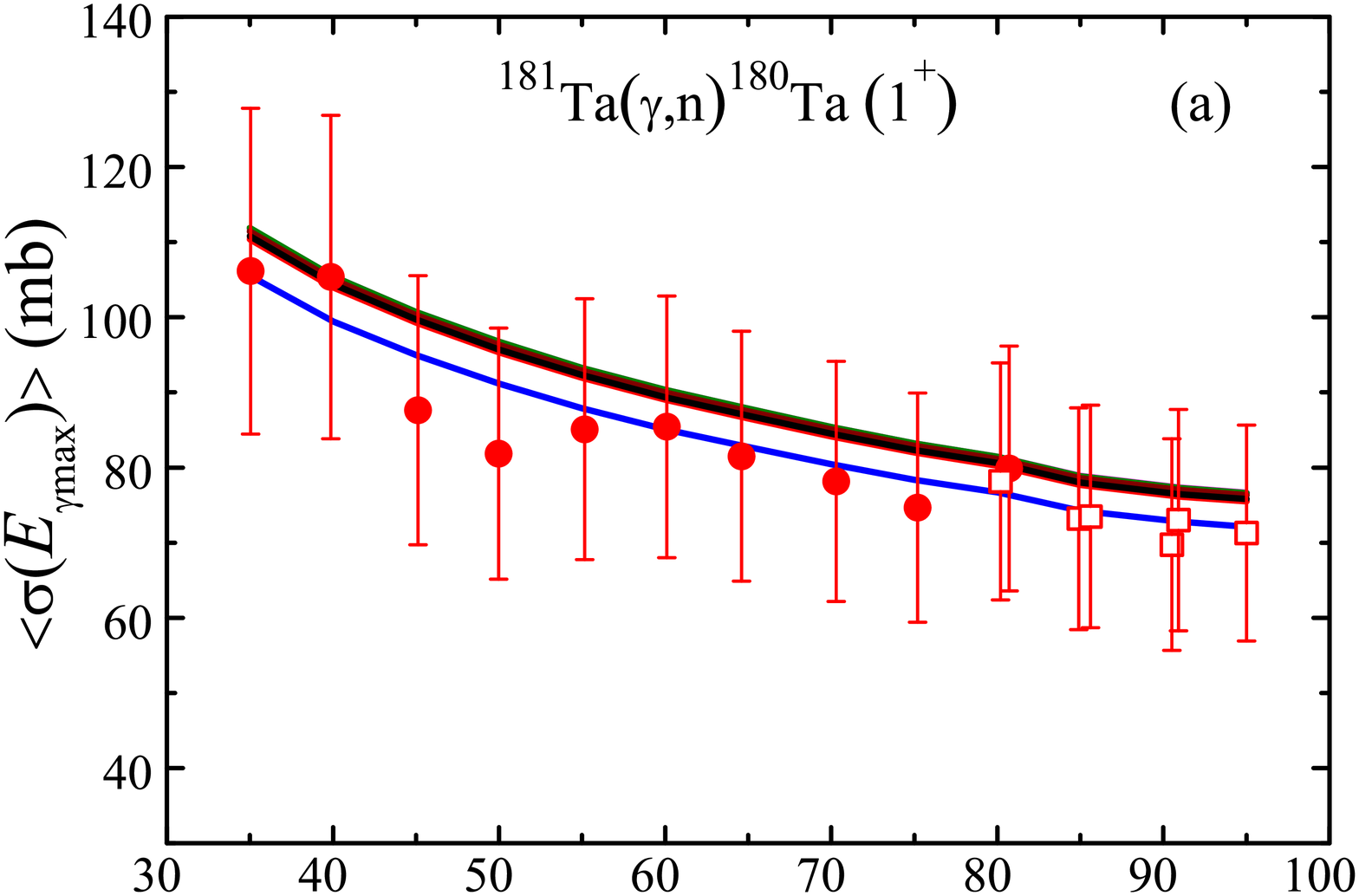}} \\
	\end{minipage}
	\hfill
	\begin{minipage}[h]{0.49\linewidth}
		\center{\includegraphics[width=1\linewidth]{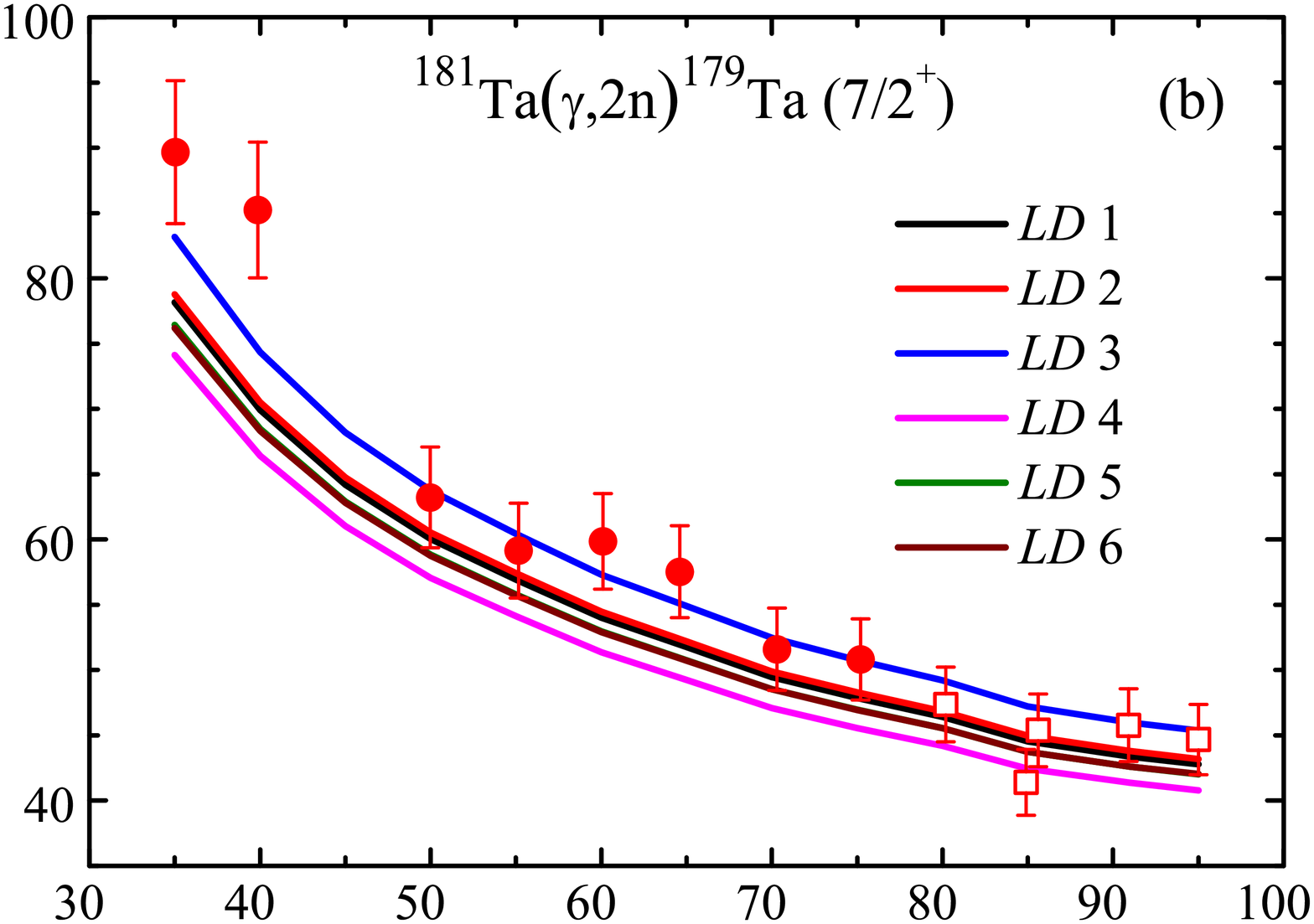}} \\
	\end{minipage}
	\vfill
	\begin{minipage}[h]{0.49\linewidth}
		\center{\includegraphics[width=1\linewidth]{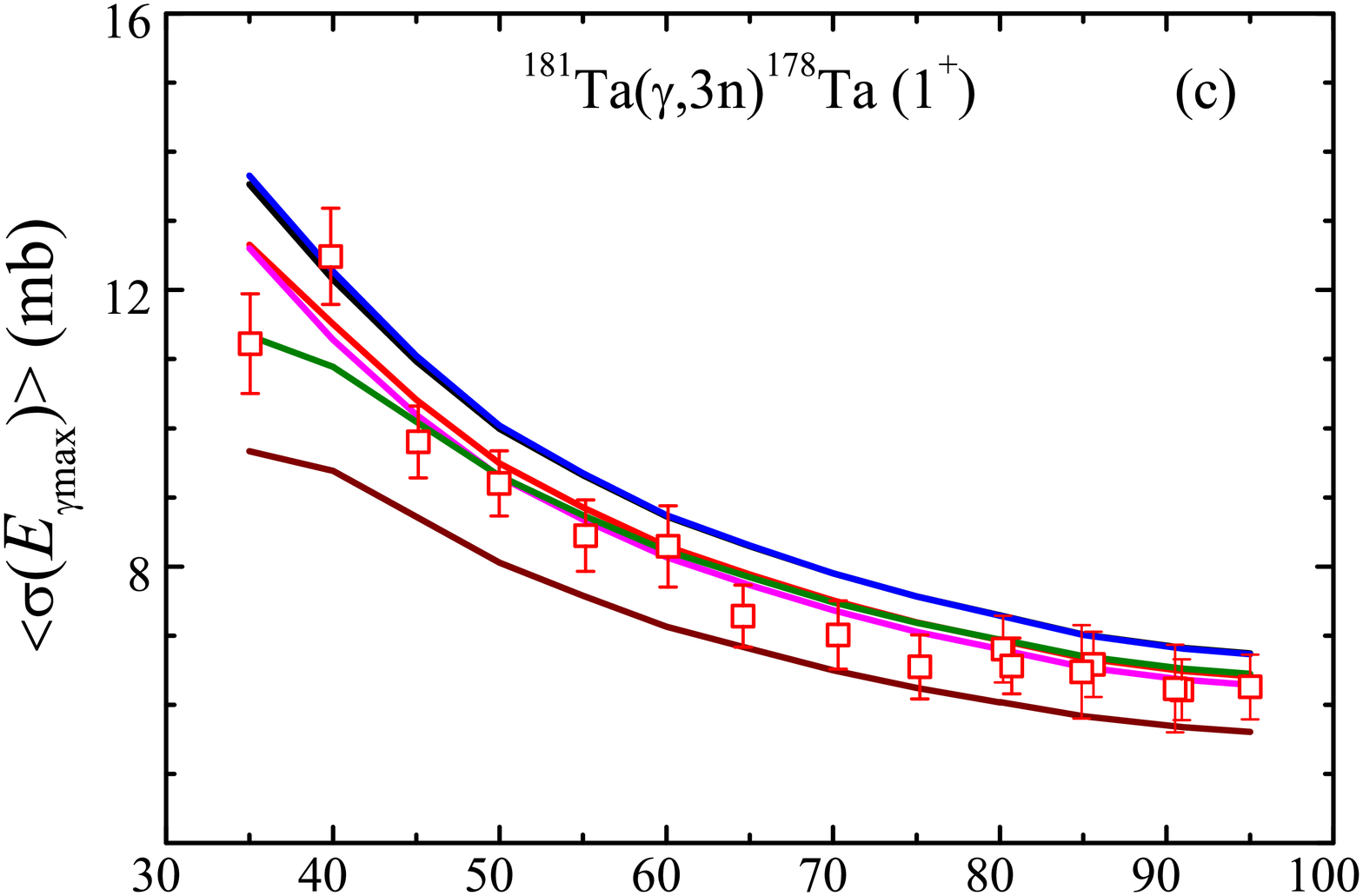}}\\
	\end{minipage}
	\hfill
	\begin{minipage}[h]{0.49\linewidth}
		\center{\includegraphics[width=1\linewidth]{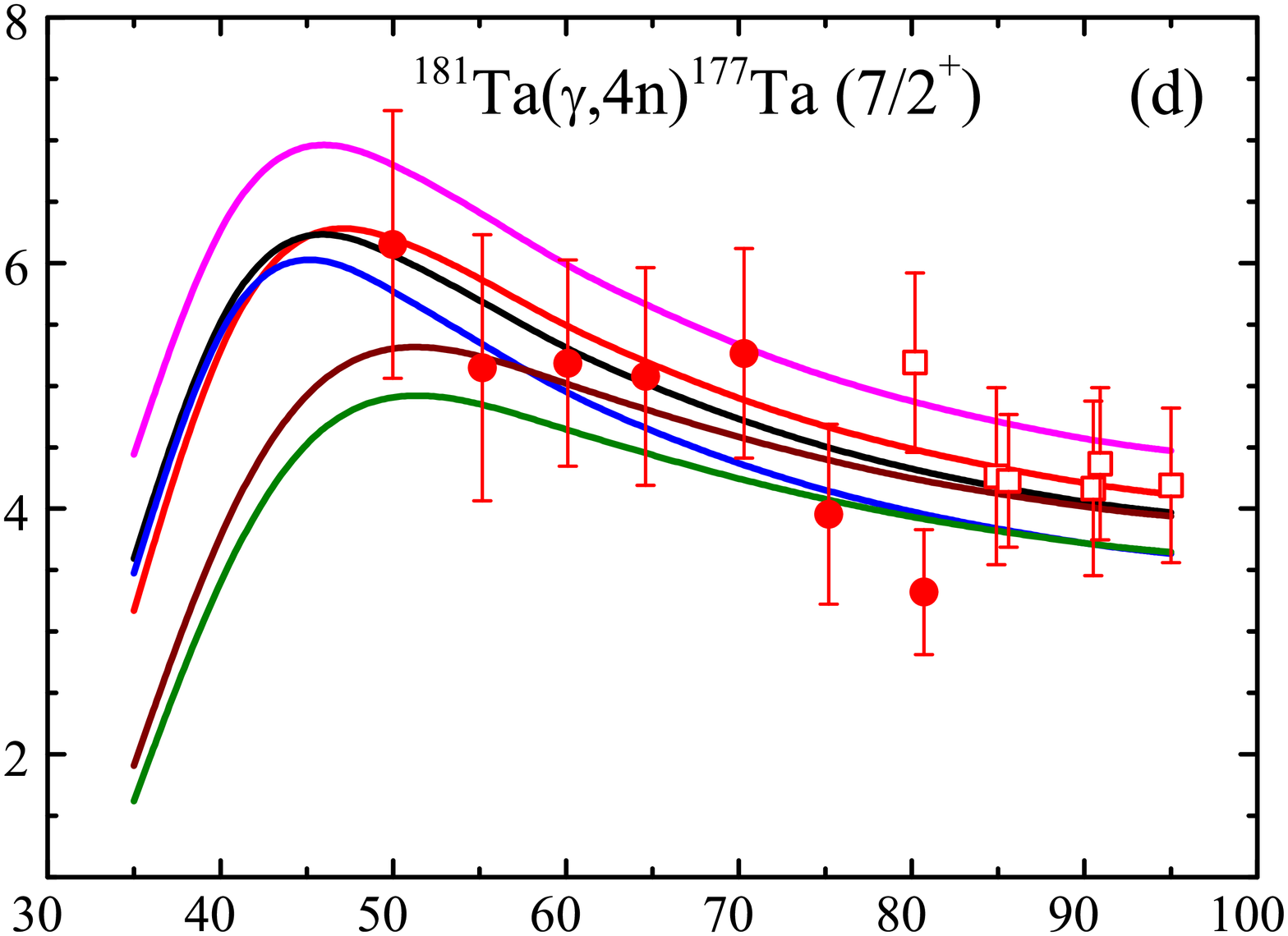}}\\
	\end{minipage}
	\begin{minipage}[h]{0.49\linewidth}
		\center{\includegraphics[width=1\linewidth]{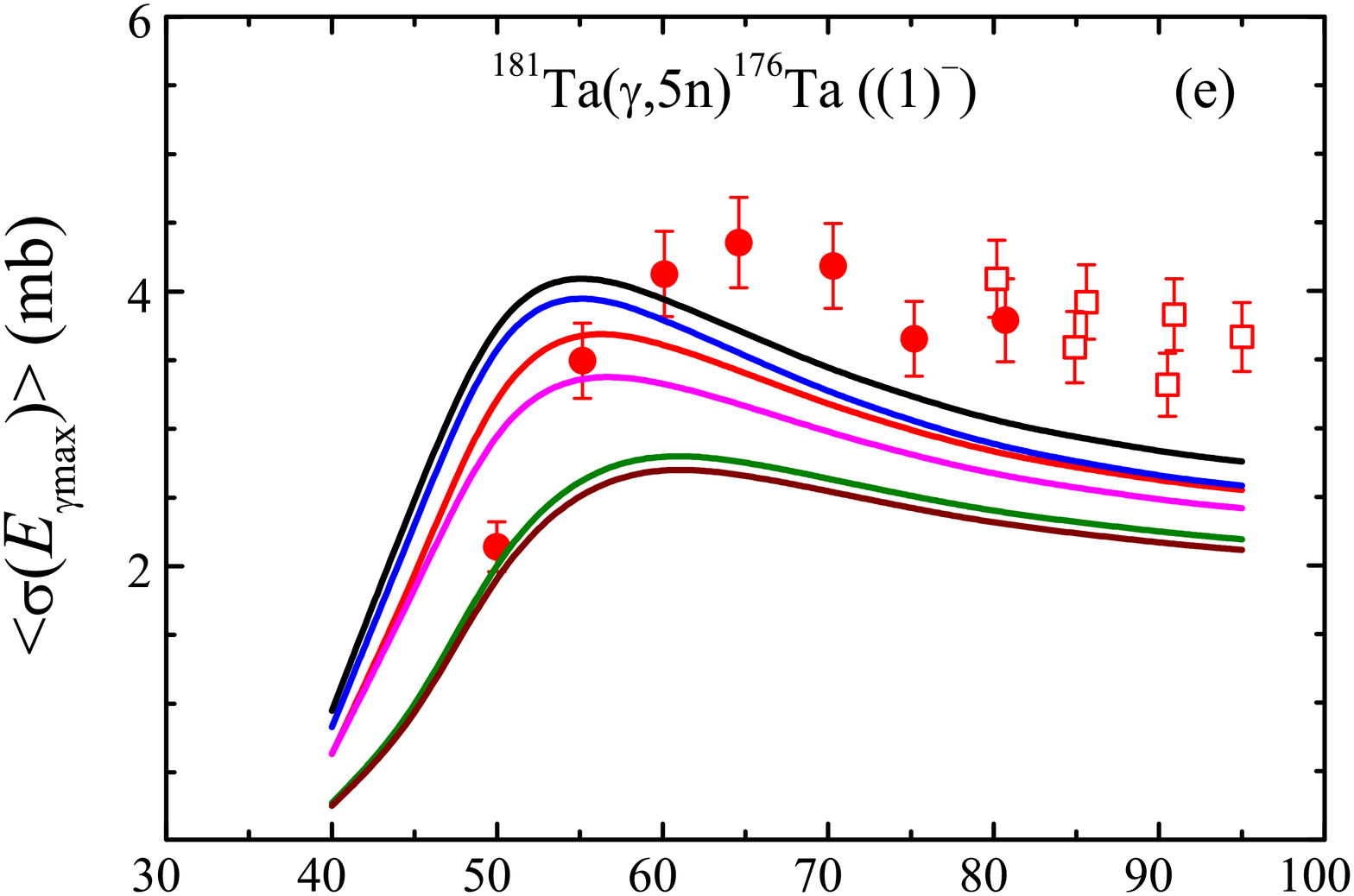}} \\
	\end{minipage}
	\hfill
	\begin{minipage}[h]{0.49\linewidth}
		\center{\includegraphics[width=1\linewidth]{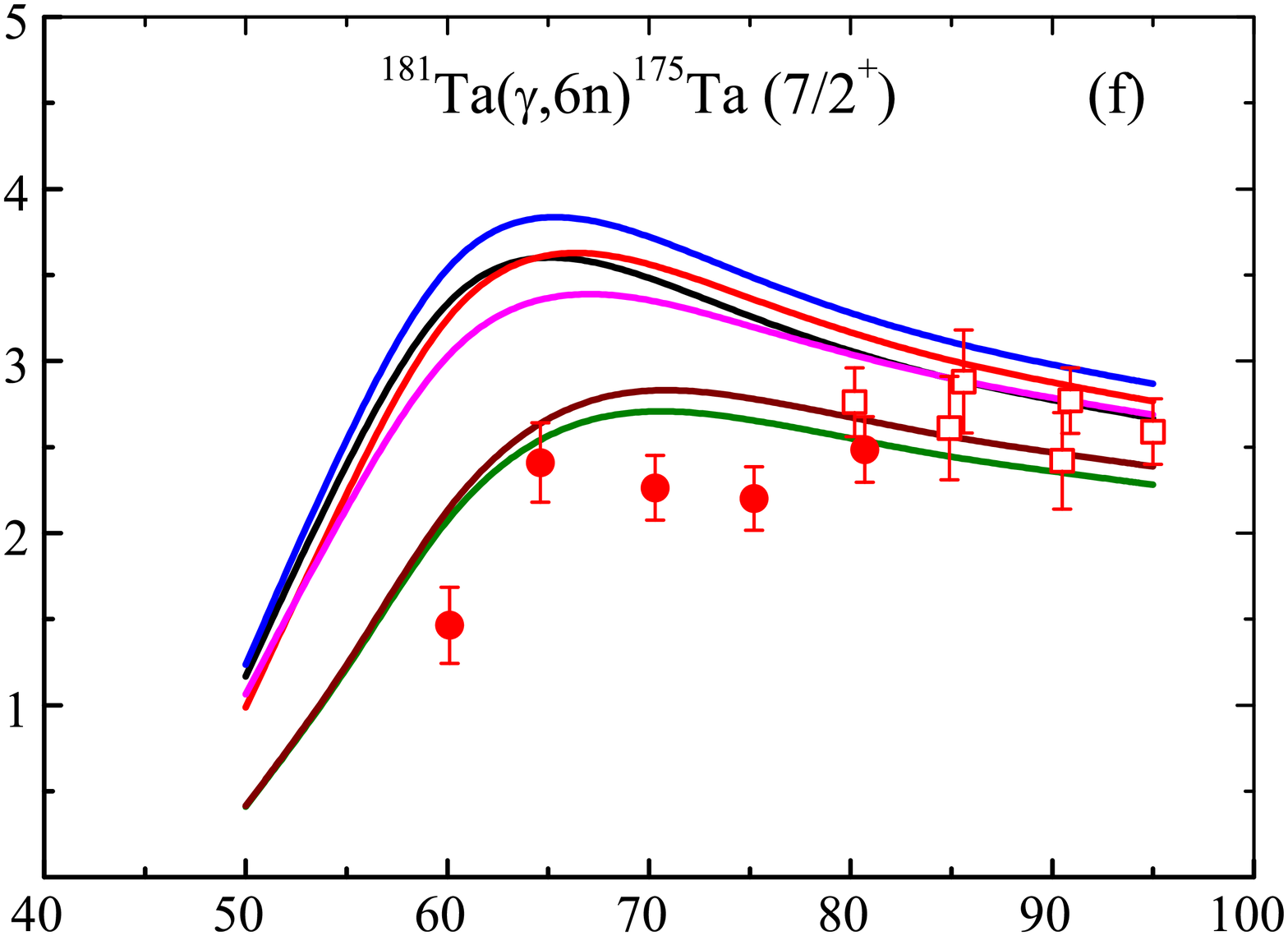}} \\
	\end{minipage}
	\vfill
	\begin{minipage}[h]{0.49\linewidth}
		\center{\includegraphics[width=1\linewidth]{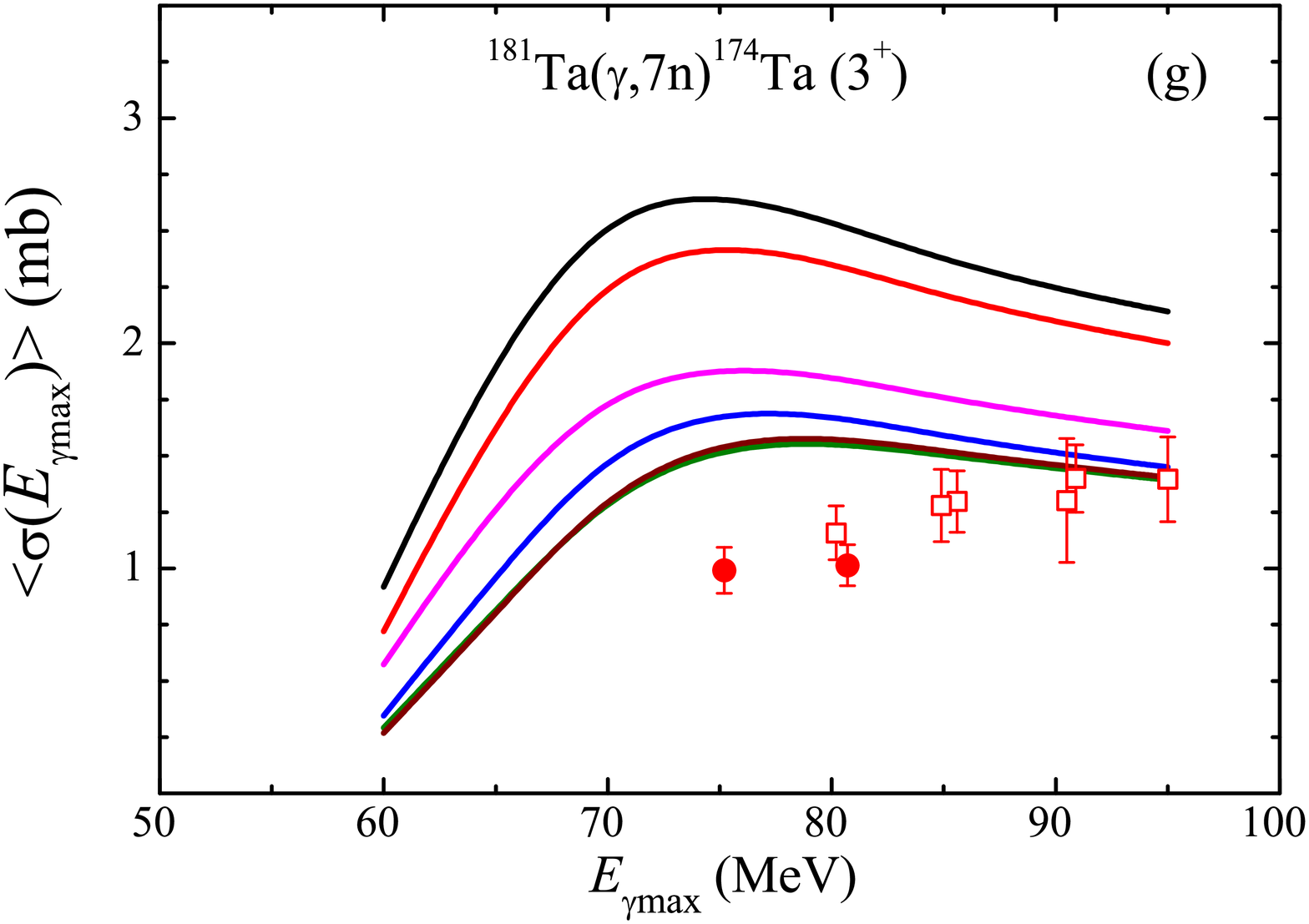}}\\
	\end{minipage}
	\hfill
	\begin{minipage}[h]{0.49\linewidth}
		\center{\includegraphics[width=1\linewidth]{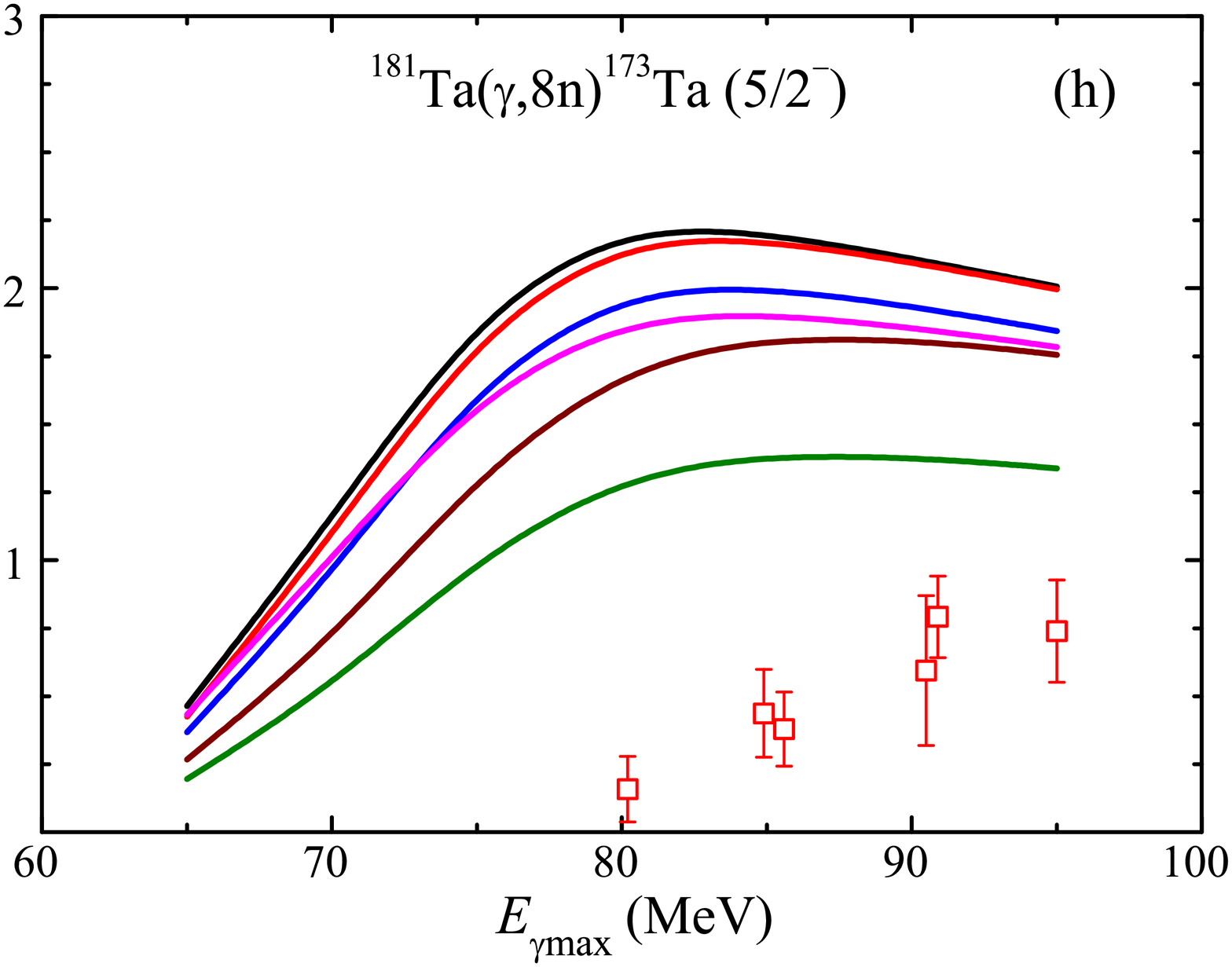}}\\
	\end{minipage}
	\caption{Total average cross-sections $\langle{\sigma(E_{\rm{\gamma max}})}\rangle$ for the reactions $^{181}\rm{Ta}(\gamma,\textit{x}n; \textit{x} \leq 8)^{181-\textit{x}}\rm{Ta}$. Full circles are our present data. Empty squares show our earlier results \cite{23}, in the case of the $^{181}\rm{Ta}(\gamma,3n)^{178}\rm{Ta}$ reaction data are taken from \cite{24}. The curves show the TALYS1.95 computations of average cross-sections using different level density models \textit{LD} 1-6.}
	\label{fig5}
\end{figure*}

The experimental and theoretical cross-sections $\langle{\sigma(E_{\rm{\gamma max}})}\rangle$ (\textit{LD} 1-6) are shown in Fig.~\ref{fig5}. For completeness, this figure shows earlier results of our studies at $E_{\rm{\gamma max}}$ = 80--95 MeV \cite{23}, and for the  $^{181}\rm{Ta}(\gamma,3n)^{178}\rm{Ta}$ reaction at $E_{\rm{\gamma max}}$ = 35--95 MeV from \cite{24}. Within the limits of experimental errors, the data of this work and from \cite{23} are in satisfactory agreement for all the reactions under consideration. Thus, the energy range of end-point bremsstrahlung $\gamma$-quantum energies $E_{\rm{\gamma max}}$  = 35--95 MeV is covered.

The cross-sections  $\langle{\sigma(E_{\rm{\gamma max}})}\rangle$ for \textit{x} = 2, 4--8 are determined in the experiment directly from the yields of the corresponding reactions. In the case of reactions $(\gamma,\rm n)$ and $(\gamma,\rm 3n)$, the procedure for determining the total cross-section is somewhat more complicated.

In the experiment for the $^{181}\rm{Ta}(\gamma,n)^{180}\rm{Ta}$ reaction, only the average cross-section 
 $\langle{\sigma(E_{\rm{\gamma max}})}\rangle_{\rm g}$ for the ground-state population of the $^{181\rm g}\rm{Ta}$ nucleus can be measured.  The total experimental average cross-section $\langle{\sigma(E_{\rm{\gamma max}})}\rangle$ for this reaction can be estimated through the use of the theoretical values of the total cross-section $\langle{\sigma(E_{\rm{\gamma max}})}\rangle$ and for the ground-state one $\langle{\sigma(E_{\rm{\gamma max}})}\rangle_{\rm g}$.  The cross-section $\langle{\sigma(E_{\rm{\gamma max}})}\rangle_{\rm g}$ / $\langle{\sigma(E_{\rm{\gamma max}})}\rangle$ ratios, computed with the TALYS1.95 code were found to be 0.915, 0.915, 0.922, 0.92, 0.875 and 0.90 for the level density models \textit{LD}~1--6, respectively. This spread in the ratios leads to uncertainty of the total cross-section estimated values of $\sim$5\%.

Note that the main contribution ($>$ 90\%) to the total cross-section comes from the cross-section for the ground state, while the addition of the metastable state is insignificant. In Fig.~\ref{fig5}(a), the experimental result is given using the value $\langle{\sigma(E_{\rm{\gamma max}})}\rangle_{\rm g}$ / $\langle{\sigma(E_{\rm{\gamma max}})}\rangle$ = 0.915. As can be seen, in the case of the reaction $(\gamma,\rm n)$, the difference between the variants of calculation by different level density models \textit{LD}~1--6 is inconsiderable. Any variant of the calculation is consistent with the experimental results within the error limits.

The experimental and calculated total cross-sections $\langle{\sigma(E_{\rm{\gamma max}})}\rangle$ for the reaction $^{181}\rm{Ta}(\gamma,2n)^{179}\rm{Ta}$ are shown in Fig.~\ref{fig5}(b). Here, too, the difference between the calculations for different models of the level density  \textit{LD}~1--6 is insignificant. As in the case of the $(\gamma,\rm n)$ reaction, it is impossible to single out the preferred model for describing the experimental results.

For the $^{181}\rm{Ta}(\gamma,3n)^{178}\rm{Ta}$ reaction, the total cross-section is determined as the sum of the experimental values of the average cross-sections of the ground $\langle{\sigma(E_{\rm{\gamma max}})}\rangle_{\rm g}$ and metastable $\langle{\sigma(E_{\rm{\gamma max}})}\rangle_{\rm m}$  states. The values $\langle{\sigma(E_{\rm{\gamma max}})}\rangle_{\rm g}$ and $\langle{\sigma(E_{\rm{\gamma max}})}\rangle_{\rm m}$ are measured directly in the experiment.

The total average cross-sections $\langle{\sigma(E_{\rm{\gamma max}})}\rangle$ for the $^{181}\rm{Ta}(\gamma,3n)^{178}\rm{Ta}$ reaction \cite{24} are shown in Fig.~\ref{fig5}{c}. The theoretical values of the total average cross-section for the reaction $(\gamma,\rm 3n)$ according to \textit{LD}1 and 6 models differ by 20--30\%, forming a corridor in which all experimental data $\langle{\sigma(E_{\rm{\gamma max}})}\rangle$ are located. The best agreement between experiment and theory is observed in the case of the \textit{LD}5 model calculation: the experimental cross-sections for the formation of a nucleus in the isomeric state $\langle{\sigma(E_{\rm{\gamma max}})}\rangle_{\rm m}$ is located below all theoretical curves, but closest to calculations using the \textit{LD}5 and 6 models, and the values $\langle{\sigma(E_{\rm{\gamma max}})}\rangle_{\rm g}$ lie above all calculations in the 35--60~MeV range, but at 65--95~MeV there is good agreement with the calculation for the \textit{LD}5 model \cite{24}.

With the increase in the number of neutrons in the reaction outlet channel the difference between the theoretical curves becomes more appreciable (Fig.~\ref{fig5}(d-h)). At first glance, it becomes possible to single out a calculation model that most optimally agrees with the experimental data. But for $x >$ 3, the reaction cross-sections decrease, and the experimental error increases, which complicates the analysis. In the case of reactions $(\gamma,\rm 4n)$ and $(\gamma,\rm 6n)$, there is a significant scatter in the experimental data, which does not allow us to choose one of the level density models. The results for $(\gamma,\rm 5n)$ and $(\gamma,\rm 8n)$ do not agree with any of the theoretical curves, although the experimental cross-sections for $(\gamma,\rm 7n)$ and $(\gamma,\rm 8n)$ are closest to the calculation by the \textit{LD}5 model.

None of the models \textit{LD} 1--6 of the TALYS1.95 code allows a satisfactory description of the entire array of experimental data obtained for the $x = $ 1--8 reactions. A joint analysis of the experimental $\langle{\sigma(E_{\rm{\gamma max}})}\rangle$ shows that, except for the reaction $(\gamma,\rm 5n)$, the most satisfactory agreement with theoretical calculations was achieved for the \textit{LD}5 model: which involves microscopic level densities (Skyrme force) from Hilaire’s combinatorial tables.

Note also the tendency to the satisfactory agreement between the experimental and calculated data on the total cross-sections $\langle{\sigma(E_{\rm{\gamma max}})}\rangle$ for photoneutron reactions on $^{181}$Ta, in which the nuclei-products are produced with positive parity $\pi$ in the ground state. For the reactions $^{181}\rm{Ta}(\gamma,3n)^{178m}\rm{Ta}$,  $^{181}\rm{Ta}(\gamma,5n)^{176}\rm{Ta}$ and $^{181}\rm{Ta}(\gamma,8n)^{173}\rm{Ta}$, which resulted in the formation of nuclei with negative parity, the experimental results do not agree with the calculations.

The numerical values of the experimental total average cross-sections $\langle{\sigma(E_{\rm{\gamma max}})}\rangle$ for the $^{181}\rm{Ta}(\gamma,\textit{x}n; \textit{x} \leq 8)^{181-\textit{x}}\rm{Ta}$ reactions in range of end-point bremsstrahlung $\gamma$-quanta energies $E_{\rm{\gamma max}}$ = 35--95 MeV are presented in Table~\ref{tab2}, \ref{tab3}, \ref{tab4}.

\begin{table}[h]
	\caption{\label{tab2} Experimental average cross-section values for the $^{181}\rm{Ta}(\gamma,\textit{x}n;\textit{x}=1, 3)^{181-\textit{x}}\rm{Ta}$ reactions}
	\centering
	\begin{ruledtabular}
		\begin{tabular}{ccccc}
			\begin{tabular}{c} Nuclear \\  reaction \end{tabular} & $E_{\rm{th}}$,~MeV & \multicolumn{3}{c}{\begin{tabular}{ccc} & $\langle{\sigma(E_{\rm{\gamma max}})}\rangle$ &  \\ 	\hline ground & meta & total \footnote{the experimental result for the total cross-section of the $^{181}\rm{Ta}(\gamma,n)^{180}\rm{Ta}$ reaction is given using the value $\langle{\sigma(E_{\rm{\gamma max}})}\rangle_{\rm g}$/$\langle{\sigma(E_{\rm{\gamma max}})}\rangle$ = 0.915; for the $^{181}\rm{Ta}(\gamma,3n)^{178}\rm{Ta}$ reaction the total cross-section is determined as the sum $\langle{\sigma(E_{\rm{\gamma max}})}\rangle_{\rm g}$ and $\langle{\sigma(E_{\rm{\gamma max}})}\rangle_{\rm m}$.}
			\end{tabular}}  \\ \hline
			              & 35.1 &  97.3$\pm$19.9  &   & 106.4$\pm$21.7   \\
$^{181}\rm{Ta}(\gamma,n)$ & 39.9 &  96.6$\pm$19.7  &   & 105.6$\pm$21.6   \\ 
			& 45.1 &  80.4$\pm$16.4  &   & 87.8$\pm$17.9  \\
			& 50.0 &  75.1$\pm$15.3  &   & 82.0$\pm$16.8  \\
			& 55.2 &  78.0$\pm$15.9  &   & 85.3$\pm$17.4  \\
			& 60.1 &  78.3$\pm$16.0  &   & 85.6$\pm$17.5  \\
			& 64.6 &  74.7$\pm$15.3  &   & 81.7$\pm$16.7  \\
			& 70.3 &  71.7$\pm$14.6  &   & 78.3$\pm$14.6  \\
			& 75.2 &  68.5$\pm$14.0  &   & 74.8$\pm$15.3  \\
			& 80.2 &  71.7$\pm$14.5  &   & 78.4$\pm$15.8  \\
			& 80.7 &  73.2$\pm$15.0  &   & 80.0$\pm$16.4  \\
			& 84.9 &  67.1$\pm$13.6  &   & 73.4$\pm$14.8  \\
			& 85.6 &  67.4$\pm$13.6  &   & 73.7$\pm$14.9  \\
			& 90.5 &  64.0$\pm$12.9  &   & 69.9$\pm$14.1  \\
			& 90.9 &  67.0$\pm$13.5  &   & 73.2$\pm$14.8  \\
			& 95.0 &  65.4$\pm$13.2  &   & 71.5$\pm$14.4  \\[4pt]  \hline
			
						   & 35.1 &  8.59$\pm$0.70    &2.63$\pm$0.17 & 11.22$\pm$0.72   \\
$^{181}\rm{Ta}(\gamma,3n)$ & 39.9 &  9.57$\pm$0.67    &2.92$\pm$0.18 & 12.49$\pm$0.69   \\ 
			& 45.1 &  7.30$\pm$0.49    &2.50$\pm$0.16 &  9.80$\pm$0.52   \\ 
			& 50.0 &  6.98$\pm$0.45    &2.23$\pm$0.14 &  9.20$\pm$0.47   \\ 
			& 55.2 &  6.33$\pm$0.50    &2.12$\pm$0.13 &  8.45$\pm$0.52   \\ 
			& 60.1 &  6.21$\pm$0.57    &2.08$\pm$0.13 &  8.29$\pm$0.59   \\ 
			& 64.6 &  5.29$\pm$0.43    &2.00$\pm$0.12 &  7.28$\pm$0.45   \\ 
			& 70.3 &  5.20$\pm$0.48    &1.81$\pm$0.11 &  7.01$\pm$0.49   \\ 
			& 75.2 &  4.86$\pm$0.45    &1.69$\pm$0.11 &  6.55$\pm$0.46   \\ 
			& 80.2 &  4.96$\pm$0.47    &1.85$\pm$0.10 &  6.81$\pm$0.48   \\ 
			& 80.7 &  4.82$\pm$0.39    &1.74$\pm$0.11 &  6.56$\pm$0.40   \\ 
			& 84.9 &  4.77$\pm$0.67    &1.71$\pm$0.09 &  6.48$\pm$0.68   \\ 
			& 85.6 &  4.78$\pm$0.46    &1.81$\pm$0.10 &  6.59$\pm$0.47   \\ 
			& 90.5 &  4.64$\pm$0.63    &1.60$\pm$0.09 &  6.24$\pm$0.63   \\ 
			& 90.9 &  4.49$\pm$0.43    &1.73$\pm$0.09 &  6.22$\pm$0.44   \\ 
			& 95.0 &  4.56$\pm$0.46    &1.70$\pm$0.09 &  6.26$\pm$0.47   \\					
		\end{tabular}
	\end{ruledtabular}	        
\end{table}

\begin{table}[h]
	\caption{\label{tab3} Experimental average cross-section values for the $^{181}\rm{Ta}(\gamma,\textit{x}n;\textit{x}=2, 4,5)^{181-\textit{x}}\rm{Ta}$ reactions}
	\centering
	\begin{ruledtabular}
		\begin{tabular}{ccc}
			\begin{tabular}{c} Nuclear \\  reaction \end{tabular} & $E_{\rm{th}}$,~MeV & $\langle{\sigma(E_{\rm{\gamma max}})}\rangle$ \\ 	\hline
			               & 35.1 &  89.7$\pm$5.5      \\
$^{181}\rm{Ta}(\gamma,2n)$ & 39.9 &  85.2$\pm$5.2    \\ 
			& 45.1 &  ---     \\ 
			& 50.0 &  63.2$\pm$3.9      \\ 
			& 55.2 &  59.2$\pm$3.6      \\ 
			& 60.1 &  59.9$\pm$3.7      \\ 
			& 64.6 &  57.5$\pm$3.5       \\ 
			& 70.3 &  51.5$\pm$3.2       \\ 
			& 75.2 &  50.8$\pm$3.1       \\ 
			& 80.2 &  47.4$\pm$2.9     \\ 
			& 80.7 &  ---           \\ 
			& 84.9 &  41.4$\pm$2.5     \\ 
			& 85.6 &  45.4$\pm$2.8     \\ 
			& 90.5 &  ---      \\ 
			& 90.9 &  45.8$\pm$2.8     \\ 
			& 95.0 &  44.7$\pm$2.7     \\ [4pt]  \hline	
			
			& 50.0 &  6.15$\pm$1.09      \\ 
			$^{181}\rm{Ta}(\gamma,4n)$  & 55.2 &  5.15$\pm$1.08      \\ 
			& 60.1 &  5.18$\pm$0.84     \\ 
			& 64.6 &  5.08$\pm$0.89       \\ 
			& 70.3 &  5.26$\pm$0.85       \\ 
			& 75.2 &  3.96$\pm$0.73       \\ 
			& 80.2 &  5.19$\pm$0.73     \\ 
			& 80.7 &  3.32$\pm$0.51     \\ 
			& 84.9 &  4.26$\pm$0.72     \\ 
			& 85.6 &  4.23$\pm$0.54   \\ 
			& 90.5 &  4.16$\pm$0.71    \\ 
			& 90.9 &  4.37$\pm$0.62   \\ 
			& 95.0 &  4.19$\pm$0.63    \\ [4pt]  \hline
			
			& 50.0 &  2.14$\pm$0.18 \\
			$^{181}\rm{Ta}(\gamma,5n)$  & 55.2 &  3.50$\pm$0.27      \\ 
			& 60.1 &  4.13$\pm$0.31     \\ 
			& 64.6 &  4.36$\pm$0.33       \\ 
			& 70.3 &  4.19$\pm$0.31       \\ 
			& 75.2 & 3.66$\pm$0.27       \\ 
			& 80.2 &  4.09$\pm$0.28     \\ 
			& 80.7 &  3.79$\pm$0.30     \\ 
			& 84.9 &  3.59$\pm$0.26     \\ 
			& 85.6 &  3.92$\pm$0.27   \\ 
			& 90.5 &  3.32$\pm$0.23    \\ 
			& 90.9 &  3.83$\pm$0.26   \\ 
			& 95.0 &  3.67$\pm$0.25    \\ 										
		\end{tabular}
	\end{ruledtabular}	        
\end{table}

\begin{table}[h]
	\caption{\label{tab4} Experimental average cross-section values for the $^{181}\rm{Ta}(\gamma,\textit{x}n;\textit{x}= 6-8)^{181-\textit{x}}\rm{Ta}$ reactions}
	\centering
	\begin{ruledtabular}
		\begin{tabular}{ccc}
			\begin{tabular}{c} Nuclear \\  reaction \end{tabular} & $E_{\rm{th}}$,~MeV & $\langle{\sigma(E_{\rm{\gamma max}})}\rangle$ \\ 	\hline
			
			& 60.1 &  1.46$\pm$0.22     \\ 
			$^{181}\rm{Ta}(\gamma,6n)$  & 64.6 &  2.41$\pm$0.23    \\ 
			& 70.3 &  2.26$\pm$0.19       \\ 
			& 75.2 &  2.20$\pm$0.18       \\ 
			& 80.2 &  2.76$\pm$0.20     \\ 
			& 80.7 &  2.49$\pm$0.19    \\ 
			& 84.9 &  2.61$\pm$0.30     \\ 
			& 85.6 &  2.88$\pm$0.30   \\ 
			& 90.5 &  2.42$\pm$0.28    \\ 
			& 90.9 &  2.77$\pm$0.19   \\ 
			& 95.0 &  2.59$\pm$0.19     \\ [4pt]  \hline						
			
			& 75.2 & 0.99$\pm$0.10       \\ 
			$^{181}\rm{Ta}(\gamma,7n)$  & 80.2 & 1.16$\pm$0.12      \\ 
			& 80.7 &  1.01$\pm$0.09    \\ 
			& 84.9 &  1.28$\pm$0.16   \\
			& 85.6 &  1.30$\pm$0.14   \\ 
			& 90.5 &  1.30$\pm$0.28    \\ 
			& 90.9 &  1.40$\pm$0.15   \\ 
			& 95.0 &  1.40$\pm$0.19     \\ [4pt]  \hline	
			
			& 80.2 & 0.16$\pm$0.12      \\ 
			$^{181}\rm{Ta}(\gamma,8n)$  & 84.9 &  0.44$\pm$0.16   \\ 
			& 85.6 &  0.38$\pm$0.14   \\
			& 90.5 &  0.60$\pm$0.28   \\ 
			& 90.9 &  0.79$\pm$0.15   \\ 
			& 95.0 &  0.74$\pm$0.19     \\ 											
		\end{tabular}
	\end{ruledtabular}	        
\end{table}

\section{\label{Conc} CONCLUSIONS}

The present work has been concerned with the study of photoneutron reactions on the $^{181}{\rm{Ta}}$ nucleus with emission of up to 8 neutrons at end-point bremsstrahlung $\gamma$-quantum energies $E_{\rm{\gamma max}}$ = 35--80 MeV with the use of the activation and off-line $\gamma$-ray spectrometric technique (residual activity technique). The experimentally determined total average cross-sections $\langle{\sigma(E_{\rm{\gamma max}})}\rangle$ of the reactions have been compared with the theoretical values computed with TALYS1.95 code for different level density models \textit{LD} 1-6. 

The analysis of the experimental $\langle{\sigma(E_{\rm{\gamma max}})}\rangle$ values has shown that the optimum agreement with the theoretical TALYS1.95-based computations has been attained for the \textit{LD}5 model, which involves microscopic level densities (Skyrme force) from Hilaire’s combinatorial tables.  

The tendency to the satisfactory agreement between the experimental and calculated data on the total cross-sections $\langle{\sigma(E_{\rm{\gamma max}})}\rangle$ for photoneutron reactions on $^{181}{\rm{Ta}}$, in which the nuclei-products are produced with positive parity $\pi$ in the ground state was observed. For the reactions $^{181}\rm{Ta}(\gamma,3n)^{178m}\rm{Ta}$,  $^{181}\rm{Ta}(\gamma,5n)^{176}\rm{Ta}$ and $^{181}\rm{Ta}(\gamma,8n)^{173}\rm{Ta}$, which resulted in the formation of nuclei with negative parity, the experimental results do not agree with any TALYS1.95 computation for the \textit{LD} 1-6 models.

The experimental data on average cross-sections $\langle{\sigma(E_{\rm{\gamma max}})}\rangle$, $\langle{\sigma(E_{\rm{\gamma max}})}\rangle_{\rm m}$, $\langle{\sigma(E_{\rm{\gamma max}})}\rangle_{\rm g}$ for the reactions in the energy range $E_{\rm{\gamma max}}$ = 35--80 MeV have been obtained for the first time. The present results extend the range of the values obtained previously at $E_{\rm{\gamma max}}$ = 80--95~MeV towards the lower-energy region, where the cross-section maxima for the reactions under study lie. 

\begin{acknowledgments}
	The authors would like to thank the staff of the linear electron accelerator LUE-40 NSC KIPT, Kharkiv, Ukraine, for their cooperation in the realization of the experiment.
\end{acknowledgments}

\section*{DECLARATION OF COMPETING INTEREST}

The authors declare that they have no known competing financial interests or personal relationships that could have appeared to influence the work reported in this paper.



\begin{thebibliography}{00}
	\bibitem{1}
	M.B. Chadwick, P. Oblozinsky, P.E. Hodgson, G. Reffo, Phys. Rev. C \textbf{44}, 814 (1991), doi.org/10.1103/PhysRevC.44.814.
	
	\bibitem{2}
	B.S. Ishkhanov and V.N. Orlin, Phys. At. Nucl. \textbf{74}, 19 (2011).
	
	\bibitem{3}
	B.S. Ishkhanov, V.N. Orlin, and S.Yu. Troschiev. Phys. At. Nucl. \textbf{75}, 253 (2012). 
	
	\bibitem{4}
	B.S. Ishkhanov, I.M. Kapitonov, A.A. Kuznetsov, V.N.~Orlin, H.D. Yoon, Photodisintegration of molybdenum isotopes  // Moscow University Physics Bulletin. Seriya 3. Fizika, Asronomiya.  \textbf{1}, 35 (2014).(In Russian)
	
	\bibitem{5} T.E. Rodrigues, J.D.T. Arruda-Neto, A. Deppman, V.P. Likhachev, J. Mesa, C. Garcia, K. Shtejer, G. Silva, S.B. Duarte, and O.A.P. Tavares, Photonuclear reactions at intermediate energies investigated via the Monte Carlo multicollisional intranuclear cascade model // Phys. Rev. C \textbf{69}, 064611 (2004), doi.org/10.1103/PhysRevC.69.064611.
	
	\bibitem{6} A. Leprêtre, H. Beil, R. Bergère, P. Carlos, J. Fagot, A.~De Miniac, and A. Veyssière, Nucl. Phys. A \textbf{367}, 237 (1981).
	
	\bibitem{7} M.B. Chadwick, P. Oblozinsky, P.E. Hodgson, G. Reffo, Phys. Rev. C \textbf{44}, 814 (1991), doi.org/10.1103/PhysRevC.44.814.
	
	\bibitem{8} M. Herman, R. Capote, B. V. Carlson, P. Oblozinsky, M.~Sin, A. Trkov, H. Wienke, V. Zerkin, “EMPIRE: Nuclear reaction model code system for data evaluation,” Nucl. Data Sheets \textbf{108}, 2655 (2007). EMPIRE - Nuclear Reaction Model Code // https://www-nds.iaea.org/empire/.
	
	\bibitem{9} A.J. Koning, S. Hilaire, and M.C. Duijvestijn, “TALYS-1.0,” EPJ Web Conf., 211 (2008). Proc. Int. Conf. on Nuclear Data for Science and Technology, 22 – 27 Apr., 2007, Nice, France, (Eds.) O. Bersillon, F. Gunsing, E. Bauge, R. Jacqmin, and S. Leray. TALYS - based evaluated nuclear data library // http://www.TALYS.eu/home/
	
	\bibitem{10} O. Iwamoto, N. Iwamoto, S. Kunieda, F. Minato, and K. Shibata, “The CCONE code system and its application to nuclear data evaluation for fission and other reactions,” Nucl. Data Sheets \textbf{131}, 259 (2016).
	
	\bibitem{11}  T. Kawano, “CoH3: The coupled-channels and HauserFeshbach code,” (2019). CNR2018: International Workshop on Compound Nucleus and Related Topics, LBNL, Berkeley, CA, USA, September 24 – 28, 2018, (Ed. J. Escher).
	
	\bibitem{12} Data Center of Photonuclear Experiments, http://cdfe.sinp.msu.ru/.
	
	\bibitem{13} Experimental Nuclear Reaction Data (EXFOR) // https://www-nds.iaea.org/exfor/.
	
	\bibitem{14} G.M. Gurevich, L.E. Lazareva, V.M. Mazur, et al., Nucl. Phys. A \textbf{351}, 257 (1981), doi.org/10.1016/0375-9474(81)90443-7.
	
	\bibitem{15} R.L. Bramblett, J.T. Caldwell, G.F. Auchampaugh and S.C. Fultz. Phys. Rev. \textbf{129}, 2723 (1963), doi.org/10.1103/PhysRev.129.2723.  
	
	\bibitem{16} R. Bergere, H. Beil and A. Veyssiere, Nucl. Phys. A \textbf{121}, 463 (1968), doi.org/10.1016/0375-9474(68)90433-8.  
	
	\bibitem{17} E.G. Fuller and M.S. Weiss, Phys. Rev. \textbf{112}, 560 (1958), doi.org/10.1103/PhysRev.112.560. 
	
	\bibitem{18} O.V. Bogdankevich, B.I. Goryachev, V.A. Zapevalov. Soviet Physics - JETP \textbf{42}, 1502 (1962). (In Russian)
	
	\bibitem{19} G.P. Antropov, I.E. Mitrofanov, B.S. Russkikh. Bulletin of the Academy of Sciences of the USSR, Physical Series \textbf{31}, 336 (1967).  (In Russian)
	
	\bibitem{20} B.S. Ishkhanov, I.M. Kapitonov, E.V. Lazutin, et al., Zhurnal Ehksp. i Teor. Fiziki, Pis’ma \textbf{10}, 80 (1969).  
	
	\bibitem{21} T. Kawano, Y.S. Cho, P. Dimitriou, et al. IAEA Photonuclear Data Library 2019 // Nuclear Data Sheets \textbf{163}, 109 (2020), doi.org/10.1016/j.nds.2019.12.002.
	
	\bibitem{22} V.A. Zheltonozhsky, M.V. Zheltonozhskaya, A.V.~Savrasov, A.P. Chernyaev, S.V.~Varzar, V.V.~Kobets, Studying the population of $^{178\rm m,177}$Ta in $(\gamma, \textit{x}\rm n)$ reactions. Phys. Part. Nuclei Lett. \textbf{18}, 315 (2021), doi.org/10.1134/S1547477121030122. 
	
	\bibitem{23} A.N. Vodin, O.S. Deiev, I.S. Timchenko, S.N. Olejnik, M.I. Ayzatskiy, V.A. Kushnir, V.V. Mytrochenko, S.A.~Perezhogin. Photoneutron reactions $^{181}\rm{Ta}(\gamma,\textit{x}n; \textit{x} = 1$--$8)^{181-\textit{x}}\rm{Ta}$ AT $E_{\rm{\gamma max}}$ = 80--95 MeV // Eur. Phys. J. A \textbf{57} 208 (2021), doi.org/10.1140/epja/s10050-021-00484-x, arXiv:2103.09859.
	
	\bibitem{24} O.S. Deiev, I.S. Timchenko, S.N. Olejnik, V.A. Kushnir, V.V. Mytrochenko, S.A. Perezhogin. Isomeric ratio of the $^{181}\rm Ta(\gamma,3n)^{178m,g}Ta$ reaction products at energy $E_{\rm{\gamma max}}$  up to 95 MeV // Chin. Phys. C 2021, doi.org/10.1088/1674-1137/ac2a95, arXiv:2108.01635.
	
	\bibitem{25} A.N. Vodin, O.S. Deiev, I.S. Timchenko, S.N. Olejnik. Cross-sections for the $^{27}\!\rm{Al}(\gamma,\textit{x})^{24}\rm{Na}$ multiparticle reaction at $E_{\gamma \rm max}$ = 40--95 MeV // Eur. Phys. J. A \textbf{57} 207 (2021), doi.org/10.1140/epja/s10050-021-00483-y, arXiv:2012.14475 
	
	\bibitem{26} A.N. Vodin, O.S. Deiev, I.S. Timchenko, S.N. Olejnik, A.S. Kachan, L.P. Korda, E.L. Kuplennikov, V.A. Kushnir, V.V. Mytrochenko, S.A. Perezhogin, N.N. Pilipenko, V.S. Trubnikov. Cross-sections for photonuclear reactions $^{93}\rm Nb(\gamma,n)^{92m}Nb$ and $^{93}\rm Nb(\gamma,n)^{92t}Nb$ at boundary energies of bremsstrahlung $\gamma$-energies $E_{\rm{\gamma max}}$  = 36…91 MeV // Probl. Atom. Sci. Tech. \textbf{3}, 148 (2020).
	
	\bibitem{27} A.N. Vodin, O.S. Deiev, V.Yu. Korda, I.S. Timchenko, S.N. Olejnik, N.I. Aizatsky, A.S. Kachan, L.P. Korda, E.L. Kuplennikov, V.A. Kushnir, V.V.~Mytrochenko, S.A. Perezhogin. Photoneutron reactions on $^{93}\rm Nb$ at $E_{\gamma \rm max}$ = 33--93 MeV // Nucl. Phys. A \textbf{1014} 122248 (2021), doi.org/10.1016/j.nuclphysa.2021.122248, arXiv:2101.08614.
	
	\bibitem{28} H. Naik, G.N. Kim, R. Schwengner, K. Kim, M. Zaman, M. Tatari, M. Sahid, S.C. Yang, R. John, R. Massarczyk, A. Junghans, S.G. Shin, Y. Key, A. Wagner, M.W. Lee, A. Goswami, M.-H. Cho, Nucl. Phys. A \textbf{916}, 168 (2013), dx.doi.org/10.1016/j.nuclphysa.2013.08.003.
	
	\bibitem{29} A.N. Dovbnya, M.I. Aizatsky, V.N. Boriskin, I.V.~Khodak, V.A. Kushnir, V.V. Mytrochenko, A.N. Opanasenko, S.A. Perezhogin, L.V. Reprintsev, A.N. Savchenko, D.L. Stepin, V.I. Tatanov, V.F. Zhiglo. Beam parameters of an S-band electron linac with beam energy of 30…100 MeV // Probl. Atom. Sci. Tech. \textbf{2}, 11 (2006).

\bibitem{30} M.I. Aizatskyi, V.I. Beloglazov, V.N. Boriskin, V.M.~Vereschaka, A.N. Vodin, R.M. Dronov, A.N.~Dovbnya, V.F. Zhiglo, I.M. Zaitsev, K.Yu.~Kramarenko, V.A.~Kushnir, V.V. Mytrochenko, A.M.~Opanasenko, S.M. Oleinik, S.A. Perezhogin, Yu.M. Ranyuk, O.O.~Repikhov, L.V. Reprintsev, D.L. Stepin, V.I.~Tatanov, V.L. Uvarov, I.V. Khodak, V.O. Tsymbal, and B.I. Shramenko, Probl. Atom. Sci. Tech. \textbf{3}, 60 (2014).

\bibitem{31} S. Agostinelli, J.R. Allison, K. Amako, et al., GEANT4 -- a simulation toolkit, Nuclear Instruments and Methods in Physics Research. A \textbf{506} (2003) 250, doi:10.1016/S0168-9002(03)01368-8, http://GEANT4.9.2.web.cern.ch/GEANT4.9.2/. 
	
	\bibitem{32} S.Y.F. Chu, L.P. Ekstrom, R.B. Firestone, The Lund/LBNL, Nuclear Data Search, Version 2.0, February 1999, WWW Table of Radioactive Isotopes, http://nucleardata.nuclear.lu.se/toi/.
	
	\bibitem{33} National Nuclear Data Center, Brookhaven National Laboratory // www.nndc.bnl.gov/nudat2/.
	
	\bibitem{34} O.M. Vodin, O.S. Deiev, S.M. Olejnik, Activation of $^{93}\rm Nb$ nuclei on the linac LUE-40 of RDC “Accelerator” and determination of photonuclear reaction cross-sections// Probl. Atom. Sci. Tech. \textbf{6}, 122 (2019).
	
	\bibitem{35} O.S. Deiev, G.L. Bochek, V.N. Dubina, S.K. Kiprich, G.P. Vasilyev, V.I. Yalovenko, V.D. Ovchinnik, M.Y.~Shulika. Bremsstrahlung of electrons and yield of neutrons from thick converters, passing gamma-radiation and neutrons through biological shielding, Probl. Atom. Sci. Tech. \textbf{3} 65 (2019) .
	
\end{thebibliography}
\end{document}